\documentclass{statsoc}
\usepackage{times}

\usepackage[a4paper,margin=1in,hmarginratio=1:1]{geometry}
\usepackage{amsmath,amssymb}
\usepackage{natbib}
\usepackage{hyperref}
\hypersetup{colorlinks=true,citecolor=blue,linkcolor=blue}

\usepackage{makecell} 
\usepackage{tikz-cd}
\usepackage{booktabs}
\usepackage{bbm}
\usepackage{url}
\usepackage{array}
\usepackage{multirow}
\usepackage{wrapfig}
\usepackage{float}
\usepackage{pdflscape}
\usepackage{tabu}
\usepackage{threeparttable}
\usepackage{threeparttablex}
\usepackage[normalem]{ulem}
\usepackage{makecell}
\usepackage{xcolor}
\usepackage{colortbl}
\usepackage{graphicx}
\usepackage{pdfpages}
\usepackage{subcaption}
\usepackage{lipsum}
\usepackage{placeins}   
\usepackage{algorithm2e}
\SetKwComment{Comment}{$\triangleright$ }{}
\RestyleAlgo{ruled}

\let\oldsection\section
\renewcommand{\section}{\FloatBarrier\oldsection}

\def\argmin{\mathop{\rm arg\,min}\limits}%
\let\hat\widehat
\let\tilde\widetilde

\def\given{{\,|\,}}
\def\ind{\ensuremath{\stackrel{\text{\tiny ind}}{\sim}}}%
\def\Pr{{\ensuremath{\mathbb P}}}%
\def\Exp{{\ensuremath{\mathbb E}}}%
\def\one{{\ensuremath{\mathbf 1}}}%

\usepackage{amsthm}
\theoremstyle{remark}
\newtheorem{example}{Example}

\title[LCGLM for OD Estimation]{Linearly Constrained Generalized Linear
Models with Applications in Origin-Destination Estimation}

\author{Likun Chou}
\address{Boston University, Boston, USA.}
\email{lilychou@bu.edu}
\author[Chou and Carvalho]{Luis Carvalho}
\address{Boston University, Boston, USA.}
\email{lecarval@bu.edu}

\begin{document}

\begin{abstract}
We show that data censored by linear constraints can be fit within the
generalized linear model (GLM) framework, recovering the expected latent
counts together with their uncertainty in a single Fisher-scoring procedure.
Our motivating application is origin–destination (OD) estimation in
transportation studies, where the latent data are the OD trip demands and
constraints are either origin and destination marginal counts per zone, as in
OD matrix estimation, or observed link counts, as in network tomography.
Casting the problem as a linearly constrained GLM lets us treat, within one
model, three features that arise routinely in practice: unmatched constraints,
trip predictors such as costs, and prior information such as diagonal
dominance and seed counts.
We demonstrate the methodology in synthetic and small scale studies and in a
larger case study based on simulated data from the BO4Mob study.
\end{abstract}


\section{Introduction}
Linear inverse problems for count data have a long tradition in Statistics
with diverse applications in many fields such as mark-recapture surveys in
ecology, inference for haplotypes in genomics, analysis of mail items for
biosecurity surveillance, and contingency table resampling.
In these problems we have \emph{latent} count data $Z$ of length $n$ but only
observe a linearly censored version of it,
\begin{equation}
\label{eq:linearconstraints}
Y = M^\top Z
\end{equation}
of length $m$, where $M$ is a $n$-by-$m$ \emph{configuration} or
\emph{censoring} matrix. The latent dimension $n$ is usually much larger than
the observed dimension $m$, turning even the reconstruction of $Z$ given $Y$
into a severely ill-posed problem and the ultimate estimation of parameters
describing the counts a challenging task.

Here we focus on two classical instances of these problems in transportation
studies: origin and destination (OD) estimation from marginal zone and link
counts. As we detail in Section~\ref{ssec:priorworks}, given the usual high
degree of indeterminacy in the latent counts $Z$, these problems traditionally
exploit heuristic solutions such as entropy maximization, iterative
proportional fitting or surrogate modeling formulations such as generalized
least square linear models with a covariance structure that mimics an ideal or
approximating model based on Poisson OD flows. More recent Bayesian approaches
have advanced ways of jointly estimating model parameters and latent counts,
but they tend to be based on Markov chain Monte Carlo (MCMC) sampling and thus
computationally complex and expensive.

We propose a new methodology based on a restricted curved exponential family
formulation~\citep{efron2022exponential,casella2024statistical} that naturally
captures the linear constraints and thus effectively marginalizes over the
latent counts. This formulation thus allows us to concentrate on parameter
estimation in a full generalized linear regression
setup~\citep{mccullagh1989generalized}, enabling us to formally derive
credible intervals and conduct hypothesis tests.

Under this setup, we exploit Bayesian Fisher scoring to efficiently obtain
estimates via optimization, avoiding MCMC sampling, and quantify uncertainty.
Moreover, adopting sparse or implicit matrix representations we are able to
computationally scale our proposed solution to large problem instances, as we
show in the context of OD estimation in Section~\ref{ssec:bo4mob}.

\subsection{Prior Work on Origin-Destination Estimation}
\label{ssec:priorworks}
Estimating an OD matrix from its margins is underspecified: many interior
tables reproduce the same row and column totals, so every method resolves the
ambiguity by imposing additional structure. Three broad families dominate the
transportation studies literature. The first selects the matrix of maximum
entropy---equivalently, minimum information relative to a prior
matrix---consistent with the observed totals
\citep{wilson1967,vanZuylenWillumsen1980}; iterative
proportional fitting (IPF), which scales an initial table to match the
margins, is the canonical algorithm in this family and has a long statistical
pedigree in contingency-table estimation
\citep{deming1940,fienberg1970,christensen2013log}. The second casts
estimation as a regression problem and
recovers the matrix by (constrained) generalized least squares
\citep{cascetta1984,bell1991}. The third embeds estimation within a
network-equilibrium assignment, recognizing that route choice depends on
congestion \citep{yang1992}. These methods are computationally efficient, but
they are posed as optimization problems and, with few exceptions, return point
estimates without a measure of uncertainty.

A parallel, model-based literature supplies that uncertainty at greater
computational cost. \citep{Maher1983} introduced Bayesian priors on the trip
matrix, and \citep{TebaldiWest1998} developed Bayesian inference for network
traffic from link counts. \citep{Hazelton1999} separated inference on the
\emph{expected} OD flows from the reconstruction of a particular realization,
proposing Bayesian estimators for each. Subsequent work used Markov chain
Monte Carlo (MCMC) to estimate time-varying matrices from sequences of counts
\citep{Hazelton2007} and extended the sampler to settings that impose no
structure on the matrix \citep{Hazelton2010}. \citep{Carvalho2013} modeled the
trip pattern as random, placed priors on the parameters governing it, and drew
posterior samples with a Gibbs sampler. These approaches deliver full
posterior uncertainty, but their cost grows rapidly with the number of zones.

\section{Linearly Constrained Models}
Suppose we have \emph{latent} \emph{discrete} data $Z$ of length $n$ but
only observe a linearly censored version of it, $Y = M^\top Z$ of length $m$.
The uncensored model is $Z_i \ind F(\xi_i, \phi)$, with $F$ a distribution in
the exponential family with canonical parameter $\xi_i$, cumulant function
$b$, and dispersion parameter $\phi$, that is,
\begin{equation}
\label{eq:fz}
\log \Pr(Z \given \xi) = \frac{Z^\top \xi - \one_n^\top\mathbf{b}(\xi)}{\phi}
+ c_Z(Z, \phi),
\end{equation}
where $\mathbf{b}(\xi) = [b(\xi_i)]_{i=1,\ldots,n}$ and $\one_n$ is the vector
of ones of length $n$. Suppose now that our model specifies a \emph{curved}
exponential family under a linear constraint
\begin{equation}
\label{eq:lc}
\xi = M \theta.
\end{equation}
Such a constraint arises naturally and, as we argue, commonly under a
canonical link and a design $X$ that spans the columns of $M$ because in this
case $\xi = X\beta \doteq M \theta$ where $\theta = B \beta$ with $B$ a
square $m$-by-$m$ full rank matrix.
Constraint~\eqref{eq:lc} allows us to infer the curved canonical parameter
$\theta$ using $Y$ only since it defines the marginal likelihood
\begin{equation}
\label{eq:fy}
\ell(\theta) \doteq \log \Pr(Y \given \theta) =
\frac{Y^\top \theta - \one_n^\top\mathbf{b}(M \theta)}{\phi} + c_Y(Y, \phi)
\end{equation}
where $c_Y(Y, \phi) = \log \sum_{Z\,:\,Y = M^\top Z} \exp\{c_Z(Z, \phi)\}$.

The marginal model~\eqref{eq:fy} is saturated for $Y$ since $\theta$ has $m$
parameters. However, there are $n$ latent observations $Z$ and we can thus
choose to model them using $p$ parameters, $m \leq p \leq n$,
in the latent GLM in~\eqref{eq:fz}. To this end, consider a $n$-by-$p$ design
matrix $X$ and the linear predictor $\eta_Z = X\beta$. Clearly, only $m$
parameters in $\beta$ are identifiable via $Y$. Let us then define $\beta =
(\alpha, \gamma)$ where $\alpha$ has $m$ \emph{essential} parameters whose
corresponding columns in $X$ span the columns of $M$ for consistency
and $\gamma$ has the remaining $p - m$ \emph{shape} parameters.
To incorporate prior information on $\beta$ let us set, for simplicity,
asymptotically conjugate independent priors
$\alpha \sim N(\alpha_0, \Lambda^{-1})$ and
$\gamma \sim N(\gamma_0, \Omega^{-1})$ where, to avoid non-identifiability,
the prior on $\gamma$ cannot be improper. In fact, the only role of $\gamma$
is to tilt the posterior space towards expected configurations for $Z$.
A few examples are helpful in understanding this setup before we discuss model
inference next.

\begin{example}
Suppose $Z_i \ind \text{\sf Po}(e^\theta)$, $i = 1, \ldots, n$, and
$Y = \sum_i Z_i$. In this case $M = \one_n$, the $n$-dimensional vector of
ones, $b(\theta) = e^{\theta}$, and so
\[
\log \Pr(Z \given \theta) =
Z^\top \one_n \theta - \one_n^\top \mathbf{b}(\one_n \theta) + c_Z(Z) =
\sum_i Z_i \theta - \one_n^\top\one_n b(\theta) + c_Z(Z).
\]
Thus $\log \Pr(Y \given \theta) = Y \theta - n e^{\theta} + c_Y(Y)$ according
to~\eqref{eq:fy}, which is expected since
$Y \sim \text{\sf Po}(n e^{\theta})$. However, $\log \Exp[Z_i] = \theta$ for
all $i$.

Suppose next that there is a special subset of indices
$J \subset \{1, \ldots, n\}$ with differentiated mean values and set
$Z_i \ind \text{\sf Po}(e^{\xi_i})$ with $\xi_i = \alpha + E_i \gamma$,
$E_i = I(i \in J)$. In this case,
\[
ne^{\theta} = \Exp[Y] = \Exp[M^\top Z] = e^\alpha \sum_i e^{E_i \gamma}
\]
and so $\theta = \alpha + \log(1 + (e^\gamma - 1)|J|/n)$. Only $\alpha$ or,
equivalently, $\theta$ are identifiable from the data, but the log-exposure
offset $\gamma$ adjusts the count intensity for the observations in $J$.
\end{example}

\begin{example}
Consider a random undirected graph on $m$ vertices with vectorized adjacency
matrix $Z$ such that, for a vertex pair $p = (i, j)$, $1 \leq i < j \leq m$,
\[
Z_p \ind \text{\sf Bern}[\text{logit}^{-1}(\theta_i + \theta_j)].
\]
In this case $n = \binom{m}{2}$ is the number of distinct pairs, and, since
for the pair $p = (i, j)$, $\xi_p = \theta_i + \theta_j$, we have that
$M_{pv} = \delta_{iv} + \delta_{jv}$ for $v = 1, \ldots, m$ and $\delta$ the
Kronecker delta. Thus, to estimate $\theta$ we just need to know
$Y = M^\top Z$, the vertex \emph{degrees}.

Given a set of vertices $J$, we can extend this ``degree-corrected'' model to
incorporate a ``community'' effect $\gamma > 0$: for a pair $p = (i, j)$,
$\text{logit}\Exp[Z_p] = \alpha_i + \alpha_j + E_{ij} \gamma$, where
$E_{ij} = I(i \in J) I(j \in J) + I(i \not\in J)I(j \not\in J)$ indicates if
$i$ and $j$ belong to the same group, both either in or out of $J$.
The $i$-th column of $M$ yields the expected degree
\[
\Exp[Y_i] =
\sum_{j \,:\, E_{ij} = 0} \text{logit}^{-1}(\alpha_i + \alpha_j) +
\sum_{j \,:\, E_{ij} = 1} \text{logit}^{-1}(\alpha_i + \alpha_j + \gamma),
\]
that is, $\gamma$ boosts the contribution to the expected degree from vertices
that belong to the same group as $i$.
\end{example}

\subsection{Bayesian Inference}
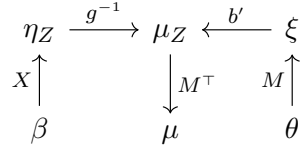
\begin{figure}
\begin{center}
\begin{tikzcd}
\eta_Z \arrow[r, "{g}^{-1}"] &
  \mu_Z \arrow[d, "M^{\top}"] &
  \xi \arrow[l, "b'"'] \\
\arrow[u, "X"] \beta & \mu & \arrow[u, "M"] \theta
\end{tikzcd}
\end{center}
\caption{Relations between model parameters.}
\label{fig:params}
\end{figure}

Under the latent GLM in~\eqref{eq:fz} we have the mean response
$\Exp[Z] \doteq \mu_Z = \mathbf{g}^{-1}(\eta_Z)$ through the link
function $g$. Moreover, still from the likelihood~\eqref{eq:fz} via the
cumulant function,
$\mu_Z = \partial(\one_n^\top \mathbf{b}(\xi))/\partial\xi =
[\partial b(\xi_i)/\partial\xi_i]_i = \mathbf{b}'(\xi)$.
Similarly, with respect to the likelihood on $Y$ we can also define
\[
\Exp[Y] \doteq \mu = \frac{\partial (\one_n^\top b(M\theta))}{\partial \theta} =
M^\top \mathbf{b}'(M\theta) = M^\top \mu_Z = \Exp[M^\top Z].
\]
Figure~\ref{fig:params} shows all relations between model parameters. The
focal point of the diagram is $\mu$, as it establishes the main relation
between $\beta$ and $\theta$:
\[
M^\top \mathbf{g}^{-1}(X\beta) = \mu = M^\top \mathbf{b}'(M\theta),
\]
that is, any $\beta$ and $\theta$ satisfying this relation are consistent with
the linear constraints encoded in $M$.

We can now obtain the maximum \emph{a posteriori} estimator for $\beta$ using
regularized Fisher scoring in an iterative reweighted least squares procedure,
as usual when fitting GLMs~\citep{mccullagh1989generalized}; the details are in
the next section, but here we derive the posterior score as it informs the
choice of priors for $\beta$.

From~\eqref{eq:fy} we have the $\theta$-score
\[
\frac{\partial\ell}{\partial\theta} =
\frac{1}{\phi}\Big(Y - M^\top \mathbf{b}'(M\theta)\Big) \doteq
\frac{1}{\phi}(Y - \mu).
\]
For the $\beta$-score we need $\partial\theta/\partial\beta =
\partial\mu/\partial\beta (\partial\mu/\partial\theta)^{-1}$, but we can use
Figure~\ref{fig:params} as a guide for the chain rules:
\[
\frac{\partial\mu}{\partial\theta} =
\frac{\partial\xi}{\partial\theta} \frac{\partial\mu_Z}{\partial\xi}
\frac{\partial\mu}{\partial\mu_Z} = M^{\top} \mathbf{V}(\mu_Z) M,
\]
where $\mathbf{V}(\mu_Z) = \text{Diag}_{i=1,\ldots,n}\{V({\mu_Z}_i)\}$ and
$V({\mu_Z}_i) = b''({b'}^{-1}({\mu_Z}_i))$ is the variance function; and
\[
\frac{\partial\mu}{\partial\beta} =
\frac{\partial\eta_Z}{\partial\beta} \frac{\partial\mu_Z}{\partial\eta_Z}
\frac{\partial\mu}{\partial\mu_Z} = X^{\top} \mathbf{N}(\mu_Z) M,
\]
where $\mathbf{N}(\mu_Z) =
\text{Diag}_{i=1,\ldots,n}\{g'({\mu_Z}_i)^{-1}\}$. Recall our prior
$\beta \sim N(\beta_0, (\Lambda \oplus \Omega)^{-1})$ with
$\beta_0 = (\alpha_0, \gamma_0)$. Now, with
$\pi(\beta) = \log\Pr(\beta\given Y)$, the posterior score is
\begin{equation}
\label{eq:score}
\begin{split}
U_\pi(\beta) := \frac{\partial\pi}{\partial\beta} &=
\frac{\partial\theta}{\partial\beta} \frac{\partial\ell}{\partial\theta} +
\frac{\partial \log\Pr(\beta)}{\partial \beta} \\
&=
\frac{1}{\phi} X^\top \mathbf{N}(\mu_Z) M (M^\top \mathbf{V}(\mu_Z) M)^{-1}
(Y - \mu) - (\Lambda \oplus \Omega)(\beta - \beta_0).
\end{split}
\end{equation}

Assume that the design matrix is partitioned according to essential and
shape parameter columns, $X = [A ~~ G]$, so that $X\beta = A\alpha + G\gamma$.
The posterior score for the essential parameters is then
\[
\frac{\partial\pi}{\partial\alpha} = \frac{1}{\phi}
A^\top \mathbf{N}(\mu_Z) M (M^\top \mathbf{V}(\mu_Z) M)^{-1}
(Y - \mu) - \Lambda(\alpha - \alpha_0).
\]
Thus, to guarantee \emph{linear consistency} for the MAP estimator
$\hat{\beta}$, that is, that the model is saturated for $Y$,
$\hat{\mu} = Y = M^\top \hat{\mu}_Z$, we need that
(i) $P \doteq A^\top \mathbf{N}(\hat{\mu}_Z) M
(M^\top \mathbf{V}(\hat{\mu}_Z) M)^{-1}$ be of full rank $m$ and (ii) that
$\Lambda = 0$, that is, an improper flat prior for $\alpha$. The first
condition is trivially satisfied if $A = MB$ for a full rank matrix $B$---that
is, if $A$ spans the columns of $M$---and $g = {b'}^{-1}$ is the canonical
link since then $\mathbf{N}(\mu_Z) = \mathbf{V}(\mu_Z)$ for all $\beta$ and so
$P = B^\top$.

\section{Bayesian Fisher Scoring}
To find the posterior mode we now resort to Newton-Raphson method or, when
approximating the Hessian by its expectation, to Fisher scoring. 
With $\Omega_\beta \doteq \Lambda \oplus \Omega$, the expected negative
Hessian is
\begin{equation}
\begin{split}
\label{eq:hessian}
H_\pi(\beta) :=
\Exp\Bigg[-\frac{\partial^2\pi}{\partial\beta \partial\beta^\top}\Bigg] &=
\frac{1}{\phi}X^\top \mathbf{N}(\mu_Z) M (M^\top \mathbf{V}(\mu_Z) M)^{-1}
\frac{\partial\mu}{\partial\beta^\top} + \Omega_\beta \\
&=
\frac{1}{\phi}X^\top \mathbf{N}(\mu_Z) M (M^\top \mathbf{V}(\mu_Z) M)^{-1}
M^\top \mathbf{N}(\mu_Z) X + \Omega_\beta.
\end{split}
\end{equation}
Now, taking the score in~\eqref{eq:score}, the update for $\beta$ at the $t$-th
iteration is then
\begin{equation}
\label{eq:fisherupdate}
\beta^{(t+1)} = \beta^{(t)} +
H_\pi\big(\beta^{(t)}\big)^{-1} U_\pi\big(\beta^{(t)}\big),
\end{equation}
or, plugging in the score and expected negative Hessian and pre-multiplying by
$H_\pi(\beta^{(t)})$,
\begin{equation*}
H_\pi\big(\beta^{(t)}\big) \beta^{(t+1)} =
\frac{1}{\phi}X^\top \mathbf{N}\big(\mu_Z^{(t)}\big) M
\big(M^\top \mathbf{V}\big(\mu_Z^{(t)}\big) M\big)^{-1}
\big[ M^\top \mathbf{N}\big(\mu_Z^{(t)}\big) \eta_Z^{(t)}
  + Y - \mu^{(t)} \big] + \Omega_\beta \beta_0,
\end{equation*}
a more familiar formulation in terms of a linear system of equations or
regularized least squares, similar to iteratively reweighted least squares.

\subsection{Initialization}
While the Fisher scoring update in~\eqref{eq:fisherupdate} is straightforward,
the main issue is to identify an initial value for $\mu_Z$
(from it $\eta_Z = g(\mu_Z)$ and $\mu = M^\top \mu_Z$ can be obtained to
compute the first update and kickstart the algorithm).
Traditionally, Fisher scoring in count GLMs rely on a perturbed
version $\mu^{(0)} \geq 0$ of the response to initialize $\mu$, such as
$\mu^{(0)} := Y + \epsilon$ with $\epsilon$ small and positive; a close,
reasonably equivalent solution here would then require solving
\begin{equation}
\label{eq:initial}
\mu_Z^{(0)} := \argmin_{\mu_Z\,:\,\mu_Z \geq 0}
\big(\mu^{(0)} - M^\top \mu_Z\big)^\top \big(\mu^{(0)} - M^\top \mu_Z\big),
\end{equation}
a quadratic program. Another approach is to use iterative proportional
fitting, which we favor since it tends to be more computationally amenable:
as we argue next, solving a quadratic program usually depends on \emph{dense}
representations (and decompositions) of $M$ which hinder computational
scalability.

\subsection{Algorithm and Computational Considerations}
\label{ssec:computational}
For the computation of the score $U_\pi$ and expected Hessian $H_\pi$ in the
update~\eqref{eq:fisherupdate} to be computationally efficient and scalable,
we need to leverage the \emph{sparseness} of the configuration matrix $M$.

If $M$ has a sparse representation we can compute a sparse QR decomposition of
\[
Q_t C_t := \mathbf{V}\big(\mu_Z^{(t)}\big)^{1/2} M,
\]
where $C_t$ is a sparse upper triangular matrix and thus the sparse Cholesky
factor of $M^\top \mathbf{V}\big(\mu_Z^{(t)}\big) M$. 
Next, we compute the Hessian factor
$X_t := C_t^{-\top} M^\top \mathbf{N}\big(\mu_Z^{(t)}\big) X$ and
the working response
\[
w_t := C_t^{-\top} M^\top \mathbf{N}\big(\mu_Z^{(t)}\big) \eta_Z^{(t)} +
Y - \mu^{(t)}
\]
so that, substituting in~\eqref{eq:hessian},
$H_\pi\big(\beta^{(t)}\big) = X_t^\top X_t/\phi + \Omega_\beta$
and so the update~\eqref{eq:fisherupdate} becomes just
\begin{equation}
\label{eq:simpleupdate}
\Bigg(\frac{1}{\phi} X_t^\top X_t + \Omega_\beta\Bigg) \beta^{(t+1)} =
\frac{1}{\phi} X_z^\top w_z + \Omega_\beta \beta_0.
\end{equation}
After convergence we obtain the maximum \emph{a posteriori} estimate
$\hat{\beta}$ and can also estimate $\hat{\text{Var}}(\beta \given Y) =
H_\pi(\hat{\beta})^{-1}$. In fact, assuming asymptotic normality in the
likelihood we have
\begin{equation}
\label{eq:asympposterior}
\beta \given Y \approx N\big(\hat{\beta}, H_\pi(\hat{\beta})^{-1}\big).
\end{equation}

If using the canonical link, $\mathbf{N} = \mathbf{V}$, and if $X = MB$ for a
full rank square matrix $B$ of order $m$, then the expected Hessian simplifies
to $H_\pi(\beta) = X^\top \mathbf{N}(\mu_Z) X / \phi + \Omega_\beta$ and
the scoring update to
\begin{equation}
\label{eq:simplifiedupdate}
\Bigg(\frac{1}{\phi} X^\top \mathbf{N}\big(\mu_Z^{(t)}\big) X +
\Omega_\beta\Bigg) \beta^{(t+1)} =
\frac{1}{\phi} \Big(
X^\top \mathbf{N}\big(\mu_Z^{(t)}\big) \eta_Z^{(t)} +
B^\top \big(Y - \mu^{(t)}\big) \Big) + \Omega_\beta \beta_0.
\end{equation}
In this case, if $\Omega_\beta$ is sparse---in fact, it is often diagonal
since it is hard to justify dependencies among parameters \emph{a
priori}---then the Hessian can be represented sparsely, admits a sparse
Cholesky decomposition and the update can be performed very efficiently.

Finally, while to assess convergence in GLM regressions we often employ
relative deviance checks, we propose to use a more robust variant based on
Pearson's $X^2$, defined as
\begin{equation}
\label{eq:pearson}
\mathbf{D}(Y, \mu_Z) := \frac{1}{\phi}\big(Y - M^\top \mu_Z\big)^\top
\big(M^\top \mathbf{V}\big(\mu_Z\big) M\big)^{-1}
\big(Y - M^\top \mu_Z\big).
\end{equation}
Algorithm~\ref{alg:scoring} below summarizes this discussion.


\begin{algorithm}
\caption{Bayesian Fisher Scoring for Linearly Constrained GLMs}
\label{alg:scoring}
\KwIn{Response counts $Y$, design matrix $X$, curved exponential family via
configuration matrix $M$ with cumulant function $b$ and link function $g$
inducing variance function $\mathbf{V}$ and inverse link derivative
$\mathbf{N}$; running parameters: maximum number of iterations $T$ and
convergence tolerance $\epsilon$}
\KwOut{Estimated coefficients $\hat{\beta}$}
\textbf{Initialize:} use IPF to define $\mu_Z^{(0)}$ directly or perturb $Y$ to
produce $\mu^{(0)}$ and then use the quadratic program in~\eqref{eq:initial};
compute $D_t = \mathbf{D}(Y; \mu_Z^{(0)})$ using~\eqref{eq:pearson}\;
\For{$t = 0, \ldots, T$}{
  Compute the sparse QR decomposition
  $Q_t C_t = \mathbf{V}\big(\mu_Z^{(t)}\big)^{1/2} M$ to obtain the sparse
  Cholesky factor $C_t$ of $M^\top \mathbf{V}\big(\mu_Z^{(t)}\big) M$\;

  Using the backsolve operator on $C_t$, compute the Hessian factor
  $X_t = C_t^{-\top} \big(M^\top \mathbf{N}\big(\mu_Z^{(t)}\big) X\big)$
  and the working response
  $w_t = C_t^{-\top} \big(M^\top \mathbf{N}\big(\mu_Z^{(t)}\big)
  \eta_Z^{(t)}\big) + Y - \mu^{(t)}$\;

  Perform a scoring update by solving~\eqref{eq:simpleupdate} to obtain
  $\beta^{(t+1)}$\;

  Compute $\eta_Z^{(t+1)} = X\beta^{(t+1)}$,
  $\mu_Z^{(t+1)} = g^{-1}(\eta_Z^{(t+1)})$, and
  $D_{t+1} = \mathbf{D}(Y, \mu_Z^{(t+1)})$\;

  \eIf{$(D_t - D_{t+1}) / D_t < \epsilon$}{
    \textbf{break} \Comment*[r]{converged}
  }{
    Set $D_t \gets D_{t+1}$\;
  }
}
\textbf{Return} $\hat{\beta} := \beta^{(t)}$
\end{algorithm}

\section{Origin-Destination Matrix Estimation}
We now apply the methodology to two classical versions of the
origin-destination (OD) estimation problem under margin and link count
(linear) constraints. In OD estimation we consider a transportation system
covering $n$ traffic analysis zones (TAZs) and wish to estimate the latent
number of trips $Z_{ij}$ between TAZ $i$ and TAZ $j$ having observed only
the margin counts
\begin{equation}
\label{eq:odmatrixconstraints0}
O_i = \sum_{j=1}^n Z_{ij}
\quad \text{and} \quad
D_j = \sum_{i=1}^n Z_{ij}
\qquad \text{for}~i, j = 1,\ldots,n.
\end{equation}
Figure~\ref{table:OD matrix} gives the structure of the OD matrix with its
margins.

\begin{figure}[htbp]
\centering
\begin{tabular}{ccccccc}
\toprule
TAZ      &    $1$   & $\cdots$ &    $j$   & $\cdots$ &    $n$   & Total \\
\midrule
$1$      & $Z_{11}$ & $\cdots$ & $Z_{1j}$ & $\cdots$ & $Z_{1n}$ & $O_1$ \\
$\vdots$ & $\vdots$ & $\ddots$ & $\vdots$ & $\ddots$ & $\vdots$ & $\vdots$ \\
$i$      & $Z_{i1}$ & $\cdots$ & $Z_{ij}$ & $\cdots$ & $Z_{in}$ & $O_i$ \\
$\vdots$ & $\vdots$ & $\ddots$ & $\vdots$ & $\ddots$ & $\vdots$ & $\vdots$ \\
$n$      & $Z_{n1}$ & $\cdots$ & $Z_{nj}$ & $\cdots$ & $Z_{nn}$ & $O_n$ \\
\midrule
Total    & $D_1$    & $\cdots$ & $D_j$    & $\cdots$ & $D_n$    & $S$ \\
\bottomrule
\end{tabular}
\caption{Origin destination matrix with latent counts $Z$ and observed margin
counts $O$ and $D$.}
\label{table:OD matrix}
\end{figure}

In a closed system that is free of observation errors we have
$\sum_i O_i = \sum_j D_j := S$ total trips and so one of the $2n$ linear
constraints in~\eqref{eq:odmatrixconstraints0} is redundant.
In this case, an equivalent set of linear constraints is
\[
S = \sum_{i=1}^n \sum_{j=1}^n Z_{ij}, \quad
O_i = \sum_{j=1}^n Z_{ij}
\quad \text{and} \quad
D_j = \sum_{i=1}^n Z_{ij},
\qquad \text{for}~i, j = 2,\ldots,n.
\]
With $O$ and $D$ the vectors with origin and destination margin counts and $Z$
the vectorized (column concatenated) OD matrix, we thus observe
$Y = (S, O^\top, D^\top)^\top = M^\top Z$ under the configuration matrix
\[
M = [
\one_n \otimes \one_n \quad
\one_n \otimes I_{n,-1} \quad
I_{n,-1} \otimes \one_n],
\]
where $\one_n$ is the vector of ones of length $n$, $I_{n,-1}$ is the
identity matrix of order $n$ with its first column removed, and $\otimes$ is
the Kronecker product. Note that $M$ is clearly binary and full rank, with
columns corresponding to the constraints on $S$, $O$, and $D$, respectively.
Moreover, while $M$ has $n^2$ rows and $2n-1$ columns, it has only 
$n(3n-2)$ non-zero entries and so its fill-in is $O(1/n)$:
when $n = 3$ the fill-in is less than 50\%, while less than 10\% when $n = 15$
and less than 1\% when $n = 150$.

To model the mean counts we can just take $X = M$ and assume Poisson flows,
\begin{equation}
\label{eq:odmodel}
\log \Exp[Z_{ij}] = \mu + \alpha_i\, I(i > 1) + \gamma_j\, I(j > 1),
\quad i, j = 1, \ldots, n,
\end{equation}
since then $\alpha_i$ and $\gamma_j$ can be naturally interpreted as origin and
destination demand effects, that is, we assume that
$\Exp[Z_{ij}] \propto A_i G_j$ with proportional factors $A_i$ and
$G_j$. While iterative proportional fitting provides an estimate for $A_i$ and
$G_j$ and thus indirectly for $\beta := (\mu, \alpha, \gamma)$, the Fisher
scoring procedure from Section~\ref{ssec:computational} can be used to further
derive estimates for $\text{Var}(\hat{\beta})$ and conduct tests based
on~\eqref{eq:asympposterior}, fully embedding OD matrix estimation within a
formal statitical setting. Note, however, that the model is saturated
with $2n-1$ constraints (observations) and parameters, so we cannot assess
model fit.

Now, because $M$ is sparse, for an arbitrary diagonal matrix $V$ the
fill-in of $M^\top V M$, symmetric of order $2n-1$, approaches 50\% as $n$
grows. Moreover, the simplified Fisher scoring update
in~\eqref{eq:simplifiedupdate} applies under a canonical link
and with $B = I_{2n-1}$. Thus, these computational simplifications, discussed
at the end of Section~\ref{ssec:computational}, allow for an efficient
procedure that fits large OD matrix estimation instances.

\subsection{Unmatched Margin Sums}
\label{ssec:external}
In practice, the margin counts commonly fail to balance,
$S_O := \sum_{i=1}^n O_i \neq \sum_{j=1}^n D_j := S_D$.
We absorb the surplus $|S_O - S_D|$ with an ``external'' zone $E$.
There are then two symmetric cases to consider:

\begin{description}
\item[$\mathbf{S_O < S_D}$:] We append one row to the OD matrix with flows
  $Z_{Ej}$ such that
  \[
  O_i = \sum_{j=1}^n Z_{ij}
  \quad \text{and} \quad
  D_j = \sum_{i=1}^n Z_{ij} + Z_{Ej}
  \qquad \text{for}~i, j = 1,\ldots,n,
  \]
  and avoid appending an external zone column by assuming $Z_{iE} = 0$ a.s.,
  $i = 1, \ldots, n, E$, for parsimony. The configuration matrix is then
  \[
  M = \Bigg[
  \one_n \otimes \begin{bmatrix} I_n \\ 0 \end{bmatrix}
  \quad I_n \otimes \one_{n+1}
  \Bigg],
  \]
  with two column sets corresponding to $O$ and $D$ constraints. We can
  proceed as in the regular OD matrix estimation case and set $X = M$, that
  is,
  \begin{equation}
  \label{eq:odmodelext0}
  \log \Exp[Z_{ij}] = \alpha_i I(i \neq E, i > 1) +
  \gamma_j I(j \neq E, j > 1), \quad i, j = 1, \ldots, n, E.
  \end{equation}
  However, we favor a centered design by setting $X = MB$ with
  \begin{equation}
  \label{eq:odextB}
  B = \Bigg[
    \frac{1}{2} \begin{bmatrix} \one_n \\ \one_n \end{bmatrix}
    \quad I_2 \otimes I_{n,-1} \quad
    \begin{bmatrix} (-1)^{I(S_O < S_D)}\one_n \\
    (-1)^{I(S_O > S_D)}\one_n \end{bmatrix}
  \Bigg],
  \end{equation}
  so that we keep the same means in~\eqref{eq:odmodel} and
  \begin{equation}
  \label{eq:odextZold}
  \log \Exp[Z_{Ej}] = \frac{\mu}{2} + \gamma_j I(j > 1) + \zeta,
  \quad j = 1, \ldots, n.
  \end{equation}
  Here the regular OD demand parameters $\mu$, $\alpha$, and $\gamma$
  retain their interpretation, and $\zeta$ is the origin demand of the
  external zone.

\item[$\mathbf{S_O > S_D}$:] Similarly, we append one column with flows
  $Z_{iE}$ such that
  \[
  O_i = \sum_{j=1}^n Z_{ij} + Z_{iE}
  \quad \text{and} \quad
  D_j = \sum_{i=1}^n Z_{ij},
  \qquad \text{for}~i, j = 1,\ldots,n,
  \]
  and set $Z_{Ej} = 0$ a.s. for $j = 1, \ldots, n, E$ to circumvent the need
  of an external zone row. The configuration matrix is now
  \[
  M = \Bigg[
  \one_{n+1} \otimes I_n \quad
  \begin{bmatrix} I_n \\ 0 \end{bmatrix} \otimes \one_n 
  \Bigg],
  \]
  again with two column sets corresponding to $O$ and $D$ constraints.
  As in the previous case, adopting $X = M$ yields the
  model in~\eqref{eq:odmodelext0}, but we set $X = MB$ with $B$ as
  in~\eqref{eq:odextB} to again keep the parameters and their
  interpretation from~\eqref{eq:odmodel}. In addition, instead
  of~\eqref{eq:odextZold} we have
  \begin{equation}
  \label{eq:odextZdlo}
  \log \Exp[Z_{iE}] = \frac{\mu}{2} + \alpha_i I(i > 1) + \zeta,
  \quad i = 1, \ldots, n,
  \end{equation}
  where $\zeta$ is now the destination demand for the external zone.
\end{description}

We note that in either case we still have a saturated model but now with $2n$
constraints due to the break in symmetry. As in the regular case, $M$ is
binary and sparse and so the same computational considerations from the
previous section apply.

\subsection{Extensions via Shape Parameters}
\label{ssec:extensions}
We now discuss three typical extensions of the basic model~\eqref{eq:odmodel}.
Because the model is saturated, these extensions are only possible if we
either observe more constraints---such as trip length distributions---to
treat the extra parameters as essential, or, otherwise, if we assume
informative priors for the now extra shape parameters. In general, we keep the
log-linear structure of~\eqref{eq:odmodel} so that the model is consistent
with the linear constraints, $\Exp[Z_{ij}] \propto A_i G_j$, but add factors
$F_{ij}$ so that now $\Exp[Z_{ij}] \propto A_i G_j F_{ij}$ for TAZs $i$ and
$j$.

\begin{description}
\item[Off-diagonal dominance:] In a transportation system with many zones
  the TAZs are more granular and so the expected number of intra-zonal trips is
  small, resulting in an off-diagonal dominant pattern in the OD matrix counts.
  We can capture this effect with
  \[
  \log \Exp[Z_{ij}] = \mu + \alpha_i I(i > 1) + \gamma_j I(j > 1) +
  \omega I(i = j),
  \qquad i, j = 1, \ldots, n,
  \]
  where $\omega$ is expected to be negative \emph{a priori} and grow in
  magnitude with $n$, say, $\omega = -\log n$ with high probability.
  In practice we can calibrate $\omega$ to reflect an expected proportion of
  intra-zonal trips.

\item[Trip costs:] Users in a transportation system make decisions based on
  trip costs, so it is only natural to incorporate these factors when modeling
  OD flows. Given costs $c_{ij}$ between TAZs $i$ and $j$, we can define a
  deterrence function $d$ so that $\Exp[Z_{ij}] \propto A_i G_j d(c_{ij})$
  with $d$ decreasing with costs.
  There are two classical functions: under a gravitational model,
  $d(c_{ij}) = c_{ij}^{-\nu}$ with $\nu > 0$, say, $\nu = 2$ a.s.;
  an exponential impedance function sets $d(c_{ij}) = \exp\{-\tau c_{ij}\}$
  with $\tau > 0$. A combined deterrence model is then
  \[
  \log \Exp[Z_{ij}] = \mu + \alpha_i I(i > 1) + \gamma_j I(j > 1)
  -\nu \log c_{ij} - \tau c_{ij},
  \qquad i, j = 1, \ldots, n.
  \]
  Calibrating $\tau$ is harder, especially in the combined model, but specific
  expected OD patterns can be checked after model fit.

\item[Seed matrices:] If there are prior counts from an OD ``seed matrix''
  $\mathbf{z}$ for the same TAZs---say, from a previous OD study---then we can
  use them to shape the latent OD flows by assuming
  $\Exp[Z_{ij}] \propto A_i G_j z_{ij}^\psi$. The shape parameter $\psi$
  controls the influence of the seed counts. The model is then
  \[
  \log \Exp[Z_{ij}] = \mu + \alpha_i I(i > 1) + \gamma_j I(j > 1) +
  \psi \log z_{ij},
  \qquad i, j = 1, \ldots, n.
  \]
  In the absence of additional constraints on the latent flows we have to set
  a prior on $\psi$. For instance, it is usual to just assume that the latent
  flows are proportional to the seed counts, that is, $\psi = 1$ a.s., and
  thus the $\log z_{ij}$ term is simply an offset.
\end{description}

\subsection{Synthetic Data Example}
We finish this section with an extended example adapted
from~\citet[Chapter~5]{ModelingTransport}.
A small city has $n = 4$ TAZs with OD pair costs and seed counts from a prior
study listed in Figure~\ref{fig:owexample} (top panels). Observed OD margin
flows and a trip length distribution (TLD) as a function of censored trip
costs are listed in the bottom panels.

\begin{figure}[htbp]
\centering
\hfill
\begin{tabular}[t]{rrrrr}
\toprule
TAZ & 1 & 2 & 3 & 4 \\
\midrule
1 & \cellcolor[HTML]{414487}{\textcolor{white}{ 3.0}} & \cellcolor[HTML]{277F8E}{\textcolor{white}{11.0}} & \cellcolor[HTML]{28AE80}{\textcolor{white}{ 18.0}} & \cellcolor[HTML]{59C864}{\textcolor{white}{22.0}} \\
2 & \cellcolor[HTML]{25858E}{\textcolor{white}{12.0}} & \cellcolor[HTML]{414487}{\textcolor{white}{ 3.0}} & \cellcolor[HTML]{25858E}{\textcolor{white}{ 12.0}} & \cellcolor[HTML]{31B57B}{\textcolor{white}{19.0}} \\
3 & \cellcolor[HTML]{1F9E89}{\textcolor{white}{15.5}} & \cellcolor[HTML]{228D8D}{\textcolor{white}{13.0}} & \cellcolor[HTML]{3A538B}{\textcolor{white}{ 5.0}} & \cellcolor[HTML]{33638D}{\textcolor{white}{ 7.0}} \\
4 & \cellcolor[HTML]{7AD151}{\textcolor{white}{24.0}} & \cellcolor[HTML]{28AE80}{\textcolor{white}{18.0}} & \cellcolor[HTML]{2F6B8E}{\textcolor{white}{ 8.0}} & \cellcolor[HTML]{3A538B}{\textcolor{white}{ 5.0}} \\
\bottomrule
\end{tabular}
\hfill
\begin{tabular}[t]{rrrrrr}
\toprule
TAZ & 1 & 2 & 3 & 4 & Total\\
\midrule
1 & \cellcolor[HTML]{414487}{\textcolor{white}{  5}} & \cellcolor[HTML]{365D8D}{\textcolor{white}{ 50}} & \cellcolor[HTML]{2A768E}{\textcolor{white}{100}} & \cellcolor[HTML]{22A785}{\textcolor{white}{200}} & 355\\
2 & \cellcolor[HTML]{365D8D}{\textcolor{white}{ 50}} & \cellcolor[HTML]{414487}{\textcolor{white}{  5}} & \cellcolor[HTML]{2A768E}{\textcolor{white}{100}} & \cellcolor[HTML]{7AD151}{\textcolor{white}{300}} & 455\\
3 & \cellcolor[HTML]{365D8D}{\textcolor{white}{ 50}} & \cellcolor[HTML]{2A768E}{\textcolor{white}{100}} & \cellcolor[HTML]{414487}{\textcolor{white}{  5}} & \cellcolor[HTML]{2A768E}{\textcolor{white}{100}} & 255\\
4 & \cellcolor[HTML]{2A768E}{\textcolor{white}{100}} & \cellcolor[HTML]{22A785}{\textcolor{white}{200}} & \cellcolor[HTML]{42BE71}{\textcolor{white}{250}} & \cellcolor[HTML]{3D4D8A}{\textcolor{white}{ 20}} & 570\\
\midrule
\addlinespace
Total & 205 & 355 & 455 & 620 & 1635\\
\bottomrule
\end{tabular}
\hfill

\medskip

\begin{tabular}[t]{lrrrrr}
\toprule
TAZ & 1 & 2 & 3 & 4 & Total \\
\midrule
$O$ & 400 & 460 & 500 & 702 & 2062 \\
$D$ & 260 & 400 & 500 & 802 & 1962 \\
\bottomrule
\end{tabular}

\medskip

\begin{tabular}[t]{lrrrrrrr}
\toprule
Cost range & (0--4] & (4--8] & (8--12] & (12--16] & (16--20] & 20+ & Total \\
\midrule
TLD        &    365 &    962 &     160 &      150 &      230 &  95 &  1962 \\
\bottomrule
\end{tabular}

\caption{Synthetic data example: OD costs $c_{ij}$ between TAZs (top left) and
OD seed counts $z_{ij}$ (top right); observed OD margin and TLD counts are
listed in the bottom panels, respectively.}
\label{fig:owexample}
\end{figure}

Because $S_O = 2062 > 1962 = S_D$, we
follow the model in Section~\ref{ssec:external} and assume external
destination flows to account for the discrepancy. The TLD provides $t = 6$
additional observations, but we lose one degree of freedom since
$\sum_{k=1}^t T_k = \sum_{i=1}^n D_j$. We then have as linear constraints,
\begin{multline}
\label{eq:owconstraints}
O_i = \sum_{j=1}^n Z_{ij} + Z_{iE}, \quad
D_j = \sum_{i=1}^n Z_{ij}, \quad \text{for}~i,j = 1,\ldots,n, \\
\text{and} \quad
T_k = \sum_{i,j\,:\,l_k < c_{ij} \leq u_k} Z_{ij}, \quad
\text{for}~k=2,\ldots,t,
\end{multline}
for an overall $2n + t - 1 = 13$ observations.

Since we have TLD counts, we can afford a more flexible model with all the
extensions in Section~\ref{ssec:extensions} besides the OD demand and external
zone effects in Section~\ref{ssec:external}. In particular, while the prior
seed counts favor off-diagonal dominance, as expected due to low intra-zonal
trip costs, we hope that the diagonal effect contributes with an additional
balancing effect. The full model is then
\[
\log \Exp[Z_{ij}] = \mu + \alpha_i I(i > 1) + \gamma_j I(j > 1) +
\omega I(i = j) - \nu \log c_{ij} - \tau c_{ij} + \psi \log z_{ij},
\tag{$M_2$}
\]
for $i, j = 1, \ldots, n$, along with the external means
in~\eqref{eq:odextZdlo}.

We compare $M_2$ to two other simpler models: $M_1$, excluding the gravity
impedance term, and $M_0$, further taking seed counts as a log offset for a
classical setup. The deviance table in Figure~\ref{fig:owdeviance} formally
compares the three models; the p-values are based on an approximate likelihood
ratio tests for $H_0:\,\psi = 1$ when taking $M_0$ as the null model relative
to $M_1$ and $H_0:\, \nu = 0$ when $M_1$ is a null model for $M_2$. While
a gravity impedance effect can be ignored at any reasonable significance
level, we keep the seed count effect and focus on $M_1$ from now on.

\begin{figure}[htbp]
\centering
\begin{tabular}{lrrrr}
\toprule
Model & Deviance & DF & $\Delta$Dev & p-value \\
\midrule
$M_0$: $O$ + $D$ + $D_E$ + Diag + Cost + offset(Seed)   &  21.14 & 3 \\
$M_1$: $O$ + $D$ + $D_E$ + Diag + Cost + Seed           &  13.51 & 2 & 7.63 & 0.01 \\
$M_2$: $O$ + $D$ + $D_E$ + Diag + Cost + Seed + Gravity &  13.22 & 1 & 0.29 & 0.59 \\
\bottomrule
\end{tabular}
\caption{Synthetic data example: analysis of deviance table.}
\label{fig:owdeviance}
\end{figure}

As Figure~\ref{fig:owcoefs} shows, the gravity impedance term has little
effect on the other coefficients in $M_2$ when compared to $M_1$ and is, in
fact, negative, failing its intended interpretation. On the other hand, the
seed count effect is so pronounced when included in $M_0$ that it seems to
render the intercept, $O_3$, and all other main $D$ margin effects not
significant in $M_1$. We, however, keep these terms to attain linear
consistency for the OD margin counts, as the estimated, expected flows in
Figure~\ref{fig:owresults} show.

\begin{figure}[htbp]
\centering
\begin{tabular}{lcrrr}
\toprule
Term          & Coefficient & \multicolumn{3}{c}{Estimate (SE)} \\
\cmidrule{3-5}
              &            &         $M_0$ &        $M_1$ &        $M_2$ \\
\midrule
$^*$Intercept & $\mu$      &   1.05 (0.23) & -0.49 (0.60) & -3.65 (2.47) \\
$^*O_2$       & $\alpha_2$ &  -0.18 (0.08) & -0.28 (0.09) & -0.22 (0.08) \\
$^*O_3$       & $\alpha_3$ &  -0.01 (0.09) & -0.04 (0.10) & -0.11 (0.15) \\
$^*O_4$       & $\alpha_4$ &  -0.61 (0.08) & -1.06 (0.19) & -1.14 (0.25) \\
$^*D_2$       & $\gamma_2$ &   0.21 (0.09) &  0.14 (0.10) &  0.17 (0.09) \\
$^*D_3$       & $\gamma_3$ &   0.14 (0.09) & -0.15 (0.15) & -0.14 (0.18) \\
$^*D_4$       & $\gamma_4$ &   0.25 (0.09) & -0.14 (0.16) & -0.28 (0.30) \\
$^*D_E$       & $\zeta$    &   2.86 (0.14) &  3.73 (0.34) &  5.33 (1.30) \\
Diagonal      & $\omega$   &   2.84 (0.19) &  3.76 (0.38) &  5.25 (1.24) \\
Cost          & $\tau$     &   0.10 (0.01) &  0.14 (0.02) &  0.23 (0.07) \\
Seed          & $\psi$     & 1.00 (offset) &  1.52 (0.19) &  1.35 (0.18) \\
Gravity       & $\nu$      &           --  &           -- & -1.96 (1.21) \\
\bottomrule
\end{tabular}
\caption{Synthetic data example: coefficient estimates and standard errors.
Starred terms correspond to OD margin effects whose columns in the design
matrix span columns in the configuration matrix and thus guarantee linear
consistency.}
\label{fig:owcoefs}
\end{figure}

As expected, the estimated OD flows from model $M_1$ seem to balance well
cost, diagonal, and seed count effects. Considering that there are no feasible
trip flows satisfying the constraints in~\eqref{eq:owconstraints}, the
estimated trip length distribution at the bottom of Figure~\ref{fig:owresults}
is in reasonable agreement with the observed TLD, with larger residuals only
in higher cost ranges.

\begin{figure}[htbp]
\centering
\begin{tabular}[t]{ccccccc}
\toprule
TAZ & 1 & 2 & 3 & 4 & E & Total\\
\midrule
1 & \cellcolor[HTML]{20938C}{\textcolor{white}{196.45}} & \cellcolor[HTML]{375A8C}{\textcolor{white}{ 55.57}} & \cellcolor[HTML]{39558C}{\textcolor{white}{ 43.63}} & \cellcolor[HTML]{34618D}{\textcolor{white}{ 71.75}} & \cellcolor[HTML]{3C508B}{\textcolor{white}{ 32.61}} & 400\\
2 & \cellcolor[HTML]{3C4F8A}{\textcolor{white}{ 31.68}} & \cellcolor[HTML]{23898E}{\textcolor{white}{170.90}} & \cellcolor[HTML]{32648E}{\textcolor{white}{ 78.06}} & \cellcolor[HTML]{26828E}{\textcolor{white}{154.69}} & \cellcolor[HTML]{3E4C8A}{\textcolor{white}{ 24.67}} & 460\\
3 & \cellcolor[HTML]{3E4C8A}{\textcolor{white}{ 24.40}} & \cellcolor[HTML]{2C738E}{\textcolor{white}{115.27}} & \cellcolor[HTML]{2A768E}{\textcolor{white}{121.89}} & \cellcolor[HTML]{1F988B}{\textcolor{white}{207.04}} & \cellcolor[HTML]{3C4F8A}{\textcolor{white}{ 31.40}} & 500\\
4 & \cellcolor[HTML]{414487}{\textcolor{white}{  7.47}} & \cellcolor[HTML]{365C8D}{\textcolor{white}{ 58.26}} & \cellcolor[HTML]{25AC82}{\textcolor{white}{256.42}} & \cellcolor[HTML]{7AD151}{\textcolor{white}{368.52}} & \cellcolor[HTML]{404688}{\textcolor{white}{ 11.33}} & 702\\
\midrule
\addlinespace
Total & 260 & 400 & 500 & 802 & 100 & 2062\\
\bottomrule
\end{tabular}

\medskip

\begin{tabular}[t]{lcccccc}
\toprule
Cost range & (1--4] & (4--8] & (8--12] & (12--16] & (16--20] & 20+ \\
\midrule
TLD & \cellcolor[HTML]{2B748E}{\textcolor{white}{365.00}} & \cellcolor[HTML]{7AD151}{\textcolor{white}{962.00}} & \cellcolor[HTML]{3C508B}{\textcolor{white}{160.00}} & \cellcolor[HTML]{3C4F8A}{\textcolor{white}{150.00}} & \cellcolor[HTML]{355E8D}{\textcolor{white}{230.00}} & \cellcolor[HTML]{414487}{\textcolor{white}{95.00}}\\
$\widehat{\text{TLD}}$ & \cellcolor[HTML]{2A778E}{\textcolor{white}{367.35}} & \cellcolor[HTML]{7AD151}{\textcolor{white}{953.87}} & \cellcolor[HTML]{3A548C}{\textcolor{white}{165.31}} & \cellcolor[HTML]{3C508B}{\textcolor{white}{139.67}} & \cellcolor[HTML]{32658E}{\textcolor{white}{256.58}} & \cellcolor[HTML]{414487}{\textcolor{white}{79.22}}\\
\addlinespace
Residual & \cellcolor[HTML]{AF305C}{\textcolor{white}{-0.12}} & \cellcolor[HTML]{C83F4B}{\textcolor{white}{0.26}} & \cellcolor[HTML]{9B2964}{\textcolor{white}{-0.41}} & \cellcolor[HTML]{E8602D}{\textcolor{white}{0.87}} & \cellcolor[HTML]{420A68}{\textcolor{white}{-1.66}} & \cellcolor[HTML]{FCA50A}{\textcolor{white}{1.77}}\\
\bottomrule
\end{tabular}
\caption{Synthetic data example: expected flows $\hat{\Exp}[Z_{ij}]$
(top panel) and TLD estimated counts $\hat{T}_k$ and Pearson residuals (bottom panel),
according to model $M_1$.}
\label{fig:owresults}
\end{figure}

\section{Network Tomography}
In network tomography~\citep{vardi1996network,TebaldiWest1998} we observe
traffic counts on specific links of a urban network with $n$ TAZs. If a
\emph{route} $r$ connects two TAZs $i$ and $j$ we then say that $r$ belongs to
the route set $\mathcal{R}_{ij}$ of the OD pair and set
$\mathcal{R} := \cup_{i, j = 1, \ldots, n} \mathcal{R}_{ij}$ as the overall
network route set. It is common to restrict each route set $\mathcal{R}_{ij}$
to a selection of small but representative routes that do not have significant
overlaps. Each route $r$ is represented by an ordered set of links
$L(r)$ connecting the origin $o(r)$ to the destination $d(r)$. Traffic flows
are then observed on links in the link set
$\mathcal{L} \subset \cup_{r \in \mathcal{R}} L(r)$. The observed link counts
$Y$ are related to the route flows $Z$ by
\begin{equation}
\label{eq:nettomoconstraints}
Y_l = \sum_{r\,:\,l \in L(r)} Z_r,
\qquad \text{for}~ l \in \mathcal{L}.
\end{equation}
The OD flows can be further obtained with
$Z_{ij} = \sum_{r\,:\,o(r) = i, d(r) = j} Z_r$, so the focus is usually on
estimating the route flows. Note, however, that route sets for OD pairs
whose origins and destinations are close in the network often contain a single
route $r$ and so $Z_r = Z_{o(r),d(r)}$. Clearly, the constraints
in~\eqref{eq:nettomoconstraints} correspond to~\eqref{eq:linearconstraints}
under the configuration matrix
\[
M = \Big[I\big(l \in L(r)\big)\Big]_{r \in \mathcal{R}, l \in \mathcal{L}}.
\]
Because $M$ is a route-link incidence matrix, it is binary and often sparse,
and so the computational advice from Section~\ref{ssec:computational} again
applies.

As in the previous section, setting $X = M$ results in the saturated and
linearly consistent model
\[
\log \Exp[Z_r] = \sum_{l\,:\, l \in L(r)} \beta_l,
\qquad r \in \mathcal{R},
\]
but users choose routes based on utilities such as travel time and comfort.
The same terms discussed in Section~\ref{ssec:extensions} can be applied here,
with the exception of (off)diagonal dominance since the zones are now network
nodes and loop routes are considered highly unusual. The baseline model is
then
\begin{equation}
\label{eq:nettomobase}
\log \Exp[Z_r] = \mu + \alpha_{o(r)} I(o(r) > 1) + \gamma_{d(r)} I(d(r) > 1),
\quad r \in \mathcal{R},
\end{equation}
with $k := 2|\mathcal{R}| - 1$ parameters. If $|\mathcal{L}| < k$  we need to
set priors on the coefficients or reduce their effective number, the rank of
the design, by, say, identifying similar zones: if nodes $i$ and $j$ have
similar OD profiles we can simply set $\alpha_i = \alpha_j$ and/or
$\gamma_i = \gamma_j$.

\begin{description}
\item[Route costs:] Given the route costs $c$, we can
  extend~\eqref{eq:nettomobase} with combined impedance terms,
  \[
  \log \Exp[Z_r] = \mu + \alpha_{o(r)} I(o(r) > 1) + \gamma_{d(r)} I(d(r) > 1)
  -\nu \log c_r - \tau c_r,
  \quad r \in \mathcal{R}.
  \]
  A more realistic model considers two levels: users traveling from node $i$
  to $j$ decide on their route from $\mathcal{R}_{ij}$ based on utilities
  (random choice level) which, in turn, depend on the route flows (e.g. travel
  times increase as the flows approach the limited capacity of links).
  The route assignment is then the stochastic user equilibrium of this bilevel
  program.
  A simpler way to incorporate this methodology is to consider an
  auto-regressive formulation where costs depend on route flows, $c_r(Z_r)$,
  but we do not elaborate on this extension here.

\item[Seed route flows:] Route flows $z$ adapted from similar networks and
  zones or taken from previous studies are also commonly used as seed counts,
  but, given the higher uncertainty in their representativity, they tend to be
  considered through a prior distribution with suitably specified precisions.
  In this case, $X = I_{|\mathcal{R}|}$ and
  $\beta \sim N(m(z), \Omega^{-1}(z))$ for some prior mean function $m$ and
  precision function $\Omega$. More importantly, when the design is the
  identity matrix we can tailor a simpler and more efficient procedure based
  on Algorithm~\ref{alg:scoring}.

  We can also extend the model~\eqref{eq:nettomobase} to account for the seed
  counts directly in the mean as a fixed effect,
  \[
  \log \Exp[Z_r] = \mu + \alpha_{o(r)} I(o(r) > 1) + \gamma_{d(r)} I(d(r) > 1)
  + \psi \log z_r,
  \quad r \in \mathcal{R}.
  \]
\end{description}

Finally, unmatched margin sums can occur, as in OD estimation, when the
configuration matrix $M$ is rank deficient but the marginal counts $Y$ are not
consistent. A simple solution is to append link count \emph{slacks} $S_l > 0$,
\begin{equation}
\label{eq:nettomoconstraintsslack}
Y_l = \sum_{r\,:\,l \in L(r)} Z_r + S_l,
\qquad \text{for~} l \in \mathcal{L}.
\end{equation}
The updated configuration matrix is then
$\tilde{M}^\top = [M^\top \, I_{|\mathcal{L}|}]$ and clearly full rank.

\subsection{Leicester Small Network Example}
\label{ssec:leicester}
\begin{figure}[htbp]
\centering
\includegraphics[scale=.125]{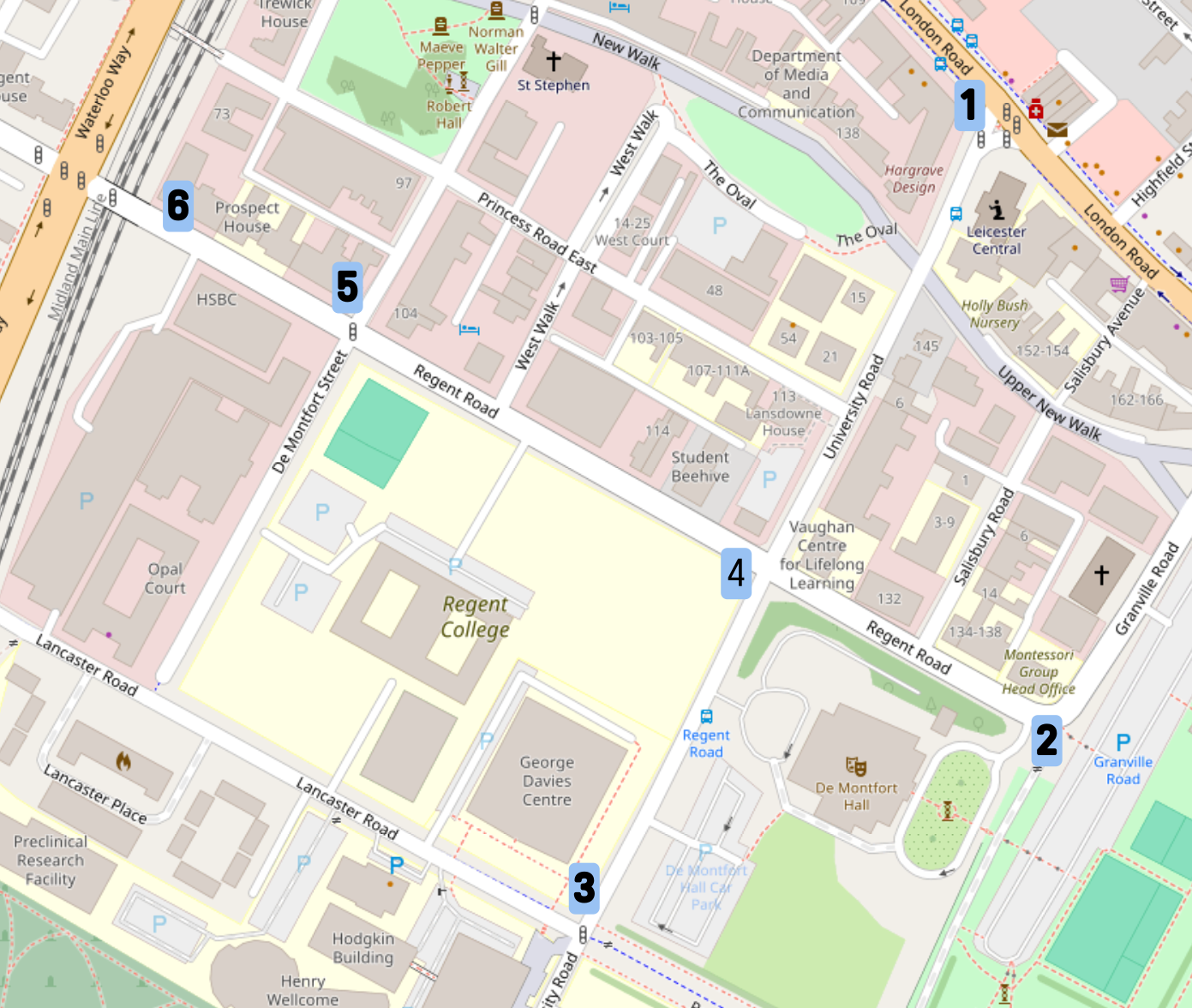}
\includegraphics[scale=.25]{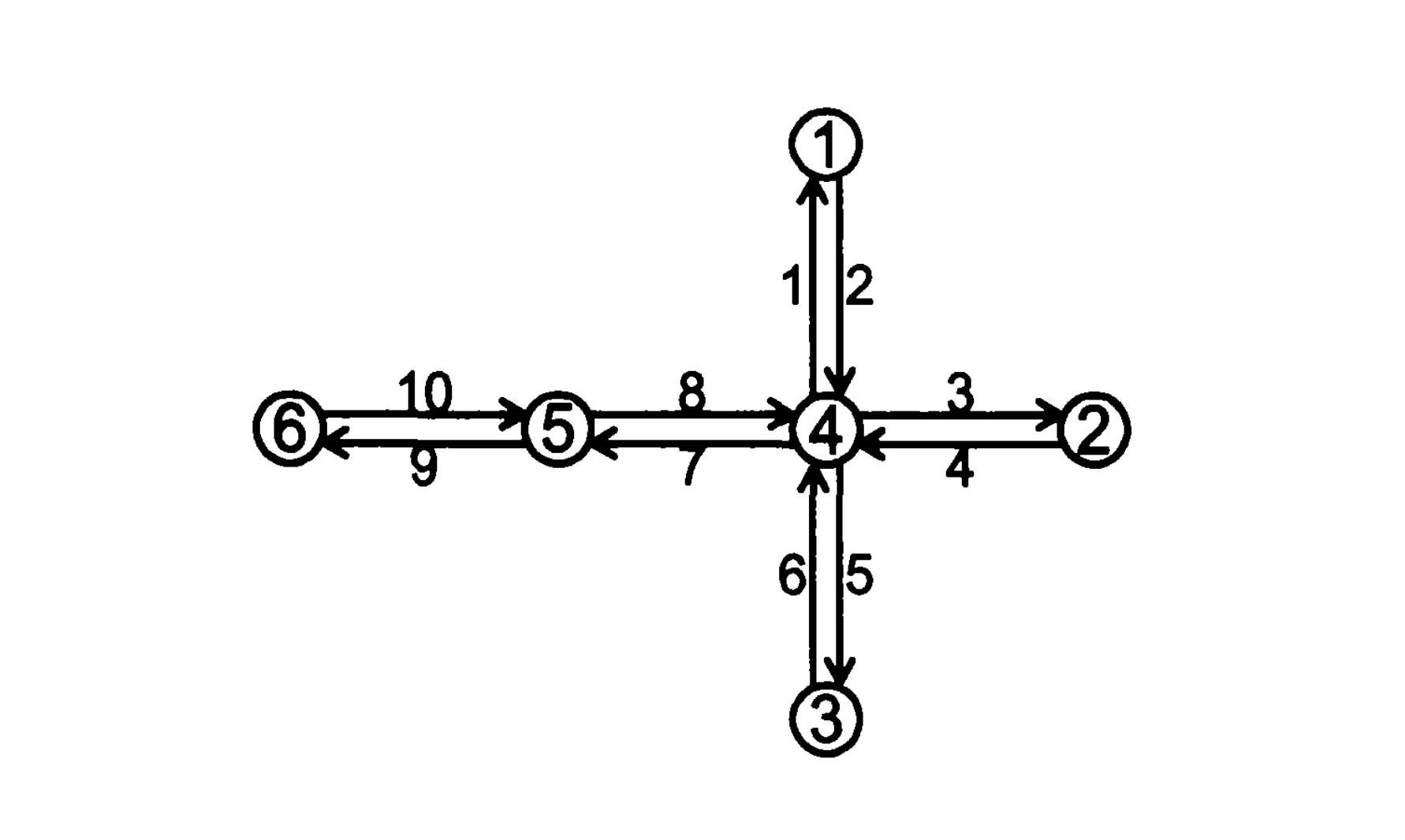}
\caption{Network tomography for small Leicester network. Left: main junctions
through Regent Rd (east-west) and University Rd (north-south); right:
schematic representation with labeled links.}
\label{fig:leicester}
\end{figure}

As an example, we estimate the route flows in a small urban network in
Leicester, UK, around the Wyggeston and Queen Elizabeth I College (former
Wyggeston Boys' school and Regent College) in the
southeast part of the city~\citep{hazelton2015network}.
The network is shown schematically in Figure~\ref{fig:leicester}.
Because the network is acyclic, every route set contains a single route, the
shortest path connecting its endpoints, and so route flow estimation is
equivalent to OD estimation. \citet{hazelton2015network} uses prior seed
counts $z$ to define a gamma prior and achieve conjugacy under a Poisson
formulation. We adopt a similar approach for comparison below, but first we
entertain a likelihood-based treatment based on route costs and seed link
counts to showcase the methodology.

The configuration matrix $M$ reported in~\citep{hazelton2015network} relates
20 OD flows (only junction 4 is not a TAZ) to link counts in 9 links (only
link 5 is not measured). In this densely measured setup, $M$ is full rank so
there is no need to append link count slack rows as
in~\eqref{eq:nettomoconstraintsslack}. Now, instead of considering a naturally
saturated model with $X = M$, we aim to incorporate route costs---unit cost
per link here, for simplicity, that is, $c_r = |L(r)|$---and the seed route
flows $z$. Because the link count \emph{net} flow between TAZs 1 and~5 are
similar, we can assume $\alpha_5 = \gamma_5 = 0$, that is, that the origin and
destination demands for TAZ~5 are the same as for TAZ~1, to gain two degrees
of freedom.

Considering the TAZ set $\mathcal{T} := \{2, 3, 6\}$, the combined, and
still saturated, full model is then
\[
\log \Exp[Z_r] = \mu + \alpha_{o(r)} I(o(r) \in \mathcal{T}) +
\gamma_{d(r)} I(d(r) \in \mathcal{T}) - \tau c_r + \psi \log z_r,
\qquad r \in \mathcal{R}.
\tag{$M_2$}
\]
We compare $M_2$ to $M_1$ under the null $H_0: \psi = 0$, that is, assuming
that the seed link counts are not informative, and to $M_0$ under the null
$H_0: \psi = \tau = 0$, assuming further that route costs are not relevant in
explaining route flows. Figure~\ref{fig:leicesterdeviance} summarizes the
model fits based on approximate likelihood ratio tests. The seed link counts
do not seem to be informative at any reasonable significance level (p-value =
0.40). While it seems that the cost term in $M_1$ is significant when compared
to $M_0$ ($H_0: \tau = 0$, p-value = 4.3 $\cdot$ 10$^{-4}$), an F-test under a
quasi-Poisson assumption yields a p-value of 0.15, not reasonably significant.

\begin{figure}[htbp]
\centering
\begin{tabular}{lrrrr}
\toprule
Model & Deviance & DF & $\Delta$Dev & p-value \\
\midrule
$M_0$: $O$ + $D$               &  13.11 & 2 \\
$M_1$: $O$ + $D$ + Cost        &   0.70 & 1 & 12.41 & 0.00 \\
$M_2$: $O$ + $D$ + Cost + Seed &   0.00 & 0 &  0.70 & 0.40 \\
\bottomrule
\end{tabular}
\caption{Leicester network example: analysis of deviance table.}
\label{fig:leicesterdeviance}
\end{figure}

Figure~\ref{fig:leicestercoefs} lists the estimated coefficients and their
standard errors for the models. Model $M_2$ has cost and seed count estimated
coefficients that are negative and with large standard errors, so $M_1$ is
preferred, as indicated by the approximate likelihood ratio test.
The main difference between $M_1$ and $M_0$ are the OD coefficients for TAZ~3;
it is not surprising that there are marginally significantly different from
zero under $M_1$ since TAZ~3 is topologically equivalent to TAZ~1 in the
network.

\begin{figure}[htbp]
\centering
\begin{tabular}{lcrrr}
\toprule
Term & Coefficient & \multicolumn{3}{c}{Estimate (SE)} \\
\cmidrule{3-5}
            &            &        $M_0$ &        $M_1$ &        $M_2$ \\
\midrule
Intercept   & $\mu$      &  1.73 (0.18) &  2.93 (0.34) &  1.62 (0.76) \\
$O_2$       & $\alpha_2$ &  1.35 (0.16) &  1.25 (0.16) &  1.99 (0.39) \\
$O_3$       & $\alpha_3$ &  0.86 (0.15) &  0.22 (0.21) &  2.72 (1.28) \\
$O_6$       & $\alpha_6$ &  1.46 (0.15) &  1.50 (0.15) &  2.95 (0.84) \\
$D_2$       & $\gamma_2$ &  1.49 (0.13) &  1.48 (0.13) &  1.90 (0.27) \\
$D_3$       & $\gamma_3$ & -0.01 (0.17) & -0.53 (0.21) &  0.76 (0.47) \\
$D_6$       & $\gamma_6$ &  0.93 (0.15) &  1.04 (0.15) &  2.11 (0.63) \\
Cost        & $\tau$     &           -- &  0.47 (0.12) & -0.44 (0.44) \\
Seed        & $\psi$     &           -- &           -- & -0.86 (0.44) \\
\bottomrule
\end{tabular}
\caption{Leicester network example: coefficient estimates and standard errors.}
\label{fig:leicestercoefs}
\end{figure}

\begin{figure}[htbp]
\centering
\includegraphics[width=.9\textwidth]{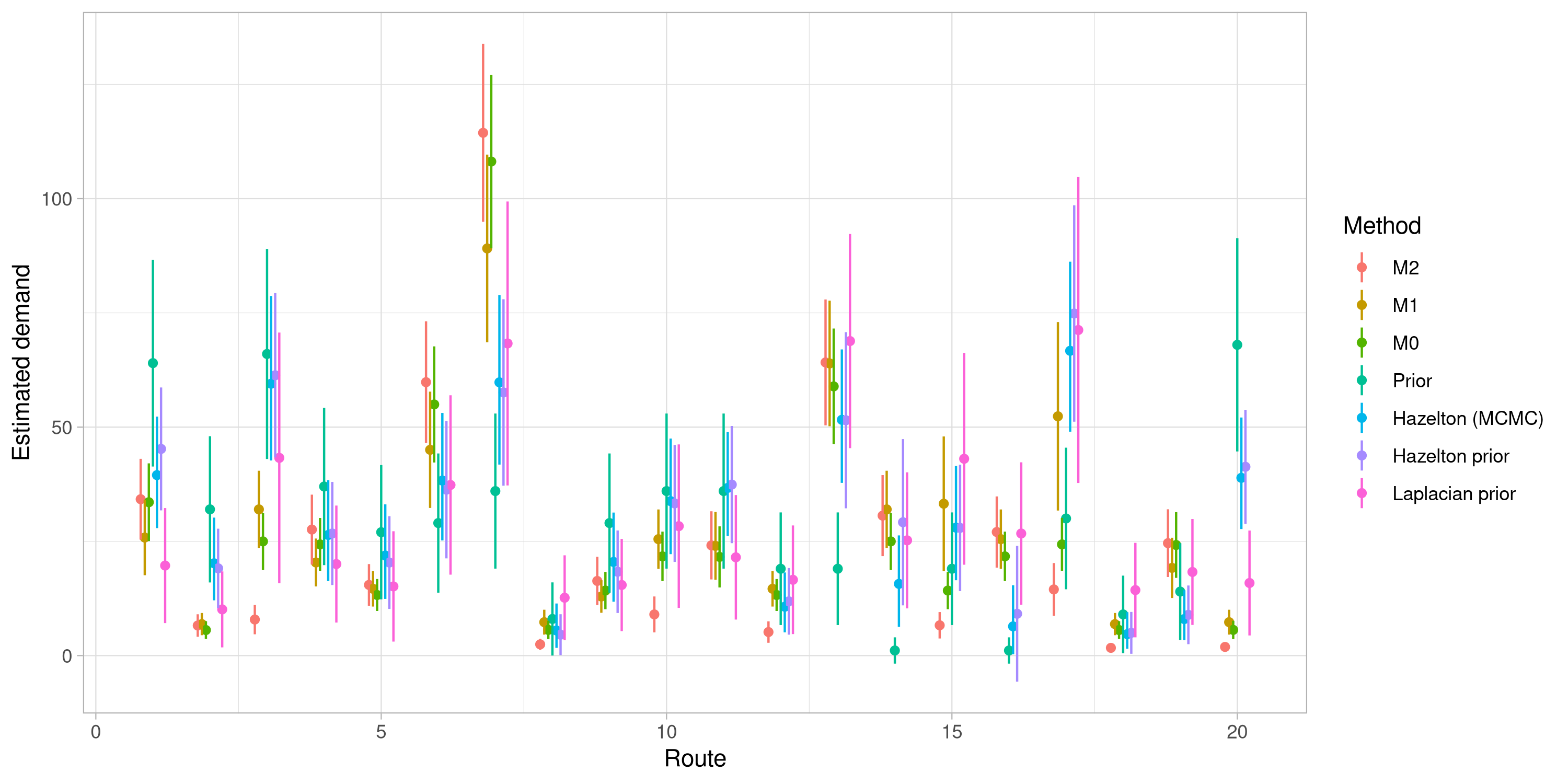}
\includegraphics[width=.9\textwidth]{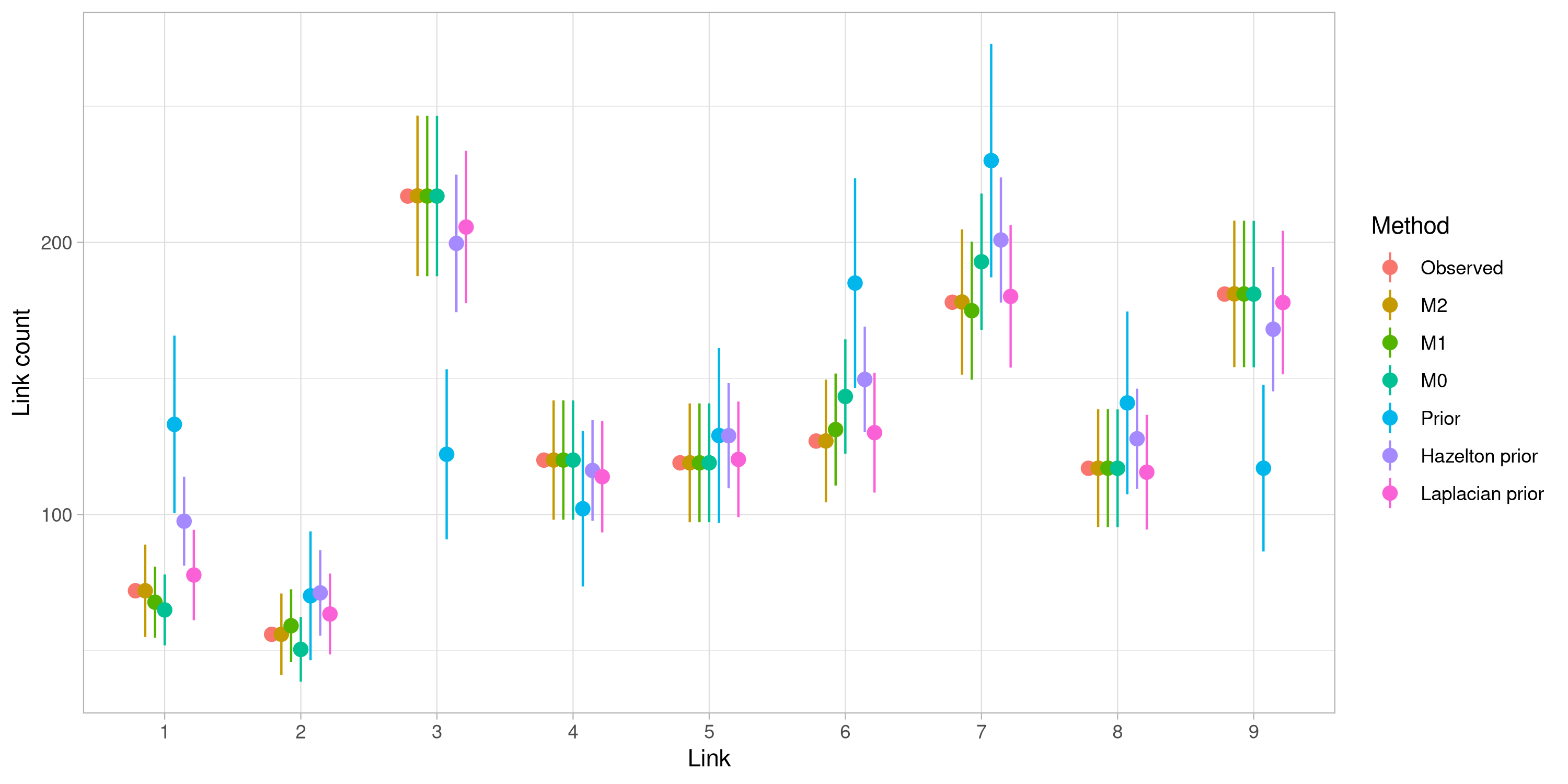}
\caption{Estimated Leicester OD demands (top) and link counts (bottom): points
represent posterior mean, lines extend to represent 95\% credible intervals.
Models $M_0$, $M_1$, and $M_2$ are likelihood-based (flat prior) using a
design based on TAZ OD demands, route cost and seed counts.
``Prior'' is based only on~\eqref{eq:leicesterprior}, not using observed link
counts. ``Hazelton (MCMC)'' lists 95\% credible intervals as reported
in~\cite{hazelton2015network}. ``Hazelton prior'' represents posterior
estimates based on the prior in~\eqref{eq:leicesterprior}.
``Laplacian prior'' shows estimates from the regularized Laplacian prior.}
\label{fig:leicesterresults}
\end{figure}

Next, we model the route demands directly by taking $X = I_{|\mathcal{R}|}$.
Because usually $|\mathcal{L}| \ll |\mathcal{R}|$, we need to set a prior
distribution for the coefficients $\beta$. \citet{hazelton2015network} uses
the prior seed counts $z$ and sets
$\beta_r \stackrel{\text{ind}}{\sim} \text{Ga}(z_r/2, 1/2)$ for conjugacy
(equivalent to assuming an identity link in our formulation). To recapitulate
his results, we keep the log link and set
\begin{equation}
\label{eq:leicesterprior}
\beta_r \stackrel{\text{ind}}{\sim}
N\big( \psi(z_r / 2) - \log(1/2), \psi_1(z_r / 2) \big),
\qquad r \in \mathcal{R},
\end{equation}
where $\psi$ and $\psi_1$ are the digamma and trigamma functions respectively,
by moment matching.
\citet{hazelton2015network} develops an MCMC procedure to estimate the flows;
we fit using Bayesian Fisher scoring and estimate the standard errors using
the Hessian; as Figure~\ref{fig:leicesterresults} shows, we obtain very
similar results, but at a fraction of the running time.

As the bottom panel of Figure~\ref{fig:leicesterresults} shows, the prior seed
counts do not represent the observed link counts well, so we adopt a different
prior. We aim now to just regularize the route flows and thus define a
weighted route graph $G$ with $\mathcal{R}$ as the vertex set and edge weights
given by the Ochiai (cosine) similarity
\begin{equation}
\label{eq:laplacianweights}
w_{rs} := \frac{|L(r) \cap L(s)|}{\sqrt{|L(r)||L(s)|}},
\qquad r, s \in \mathcal{R}, r \neq s,
\end{equation}
and $w_{rr} = 0$ for $r \in \mathcal{R}$.
The prior is then $\beta \sim N\big(0, \lambda^{-1}\Omega_G^-\big)$, where
\[
\Omega_G := \text{Diag}_{r \in \mathcal{R}}\Bigg\{\sum_{s \neq r} w_{rs}\Bigg\}
- \big[w_{rs}\big]_{r, s \in \mathcal{R}}
\]
is the weighted Laplacian of $G$. We note that $\Omega_G$ has
$\one_{\mathcal{R}}$ as its single null eigenvector and so it has rank
$|\mathcal{R}| - 1$. We determine the hyper parameter $\lambda$, the
regularization penalty, by minimizing leave-one-out cross-validation, setting
$\lambda = 1.5$.

Overall, the OD demands show good agreement between the estimates from models
$M_0$, $M_1$, and $M_2$ and the regularized Laplacian prior model, as seen in
the top panel of Figure~\ref{fig:leicesterresults}. The main exceptions are
three routes in the west-to-east direction of Regent Rd: route~3 (TAZs~5
to~2), route~7 (TAZs~6 to~2), and route~17 (TAZs~6 to~5). The credible
intervals for the oversaturated models are much larger, reflecting not only
uncertainty from weakly informative priors but potential multimodality in the
posterior space of route flows. The bottom panel offers a linear consistency
check between posterior estimates and observed link counts: $M_2$ is
consistent, $M_1$ and $M_0$ close to consistent, and the regularized Laplacian
model less linearly consistent, but with 95\% credible intervals still
covering the observed counts. Interestingly, the Hazelton seed count model is
reasonably consistent even though the prior counts are not very
representative.

\subsection{Case Study: BO4Mob}
\label{ssec:bo4mob}
In the BO4Mob benchmark framework~\citep{bo4mob}, link measurement data for a
real-world network in San Jose, CA is simulated based on actual observations
from Caltrans Performance Measurement System (PeMS). Here we consider the
\texttt{4smallRegion} network. After removing routes and OD pairs that do not
cover any measured links, we have $\mathcal{R} =$ 258 routes covering 150 OD
pairs. For simplicity of exposition, we focus on the weekdays of the first
week in the dataset, from October 10th (Monday) to October 14th 2022 (Friday).
Link counts are observed in three time periods: early morning (6--7AM), peak
morning (8--9AM), and peak afternoon (5--6PM).
There are then $D = 5$ date levels, $T = 3$ time periods, and overall
$D \cdot T = 15$ observations. There are $|\mathcal{L}| = 41$ measurement
links in the network.

The configuration matrix $M_0$ for each date-time period is rank defficient,
so we append link count slacks $S$ to set the full configuration matrix as
\[
M = \begin{bmatrix}
  I_{DT} \otimes M_0 \\ I_{DT} \otimes I_{|\mathcal{L}|}
\end{bmatrix},
\]
with route row blocks indexed first by date and then time period and similarly
for the link slack blocks. The link slack block guarantees that $M$ has full
rank $DT|\mathcal{L}|$. With the natural design $X = M$ we have the saturated
model
\[
\log \Exp[Z_{rdt}] = \sum_{l \in L(r)} \mu_{ldt}
\quad \text{and} \quad
\log \Exp[S_{ldt}] = \mu_{ldt}, \tag{$M_s$}
\]
for the mean route flows $Z_{rdt}$ and link count slacks $S_{ldt}$,
for all routes $r \in \mathcal{R}$, links $l \in \mathcal{L}$,
dates $d =1, \ldots, D$, and time periods $t = 1, \ldots, T$.

Our main goal is to find simpler patterns in the OD demands, so, as in the
Leicester network example, we start with a likelihood-based (flat prior)
approach. Our first attempt is to remove date and time interactions for each
link in $M_s$,
\[
\begin{split}
  \log \Exp[Z_{rdt}] &= \sum_{l \in L(r)} \mu_l +
  \delta_{dl} I(d > 1) + \tau_{tl} I(t > 1) \qquad \text{and} \\
  \log \Exp[S_{ldt}] &= \mu_l +
  \delta_{dl} I(d > 1) + \tau_{tl} I(t > 1).
\end{split} \tag{$M_2$}
\]

\begin{figure}
\centering
\includegraphics[width=.9\textwidth]{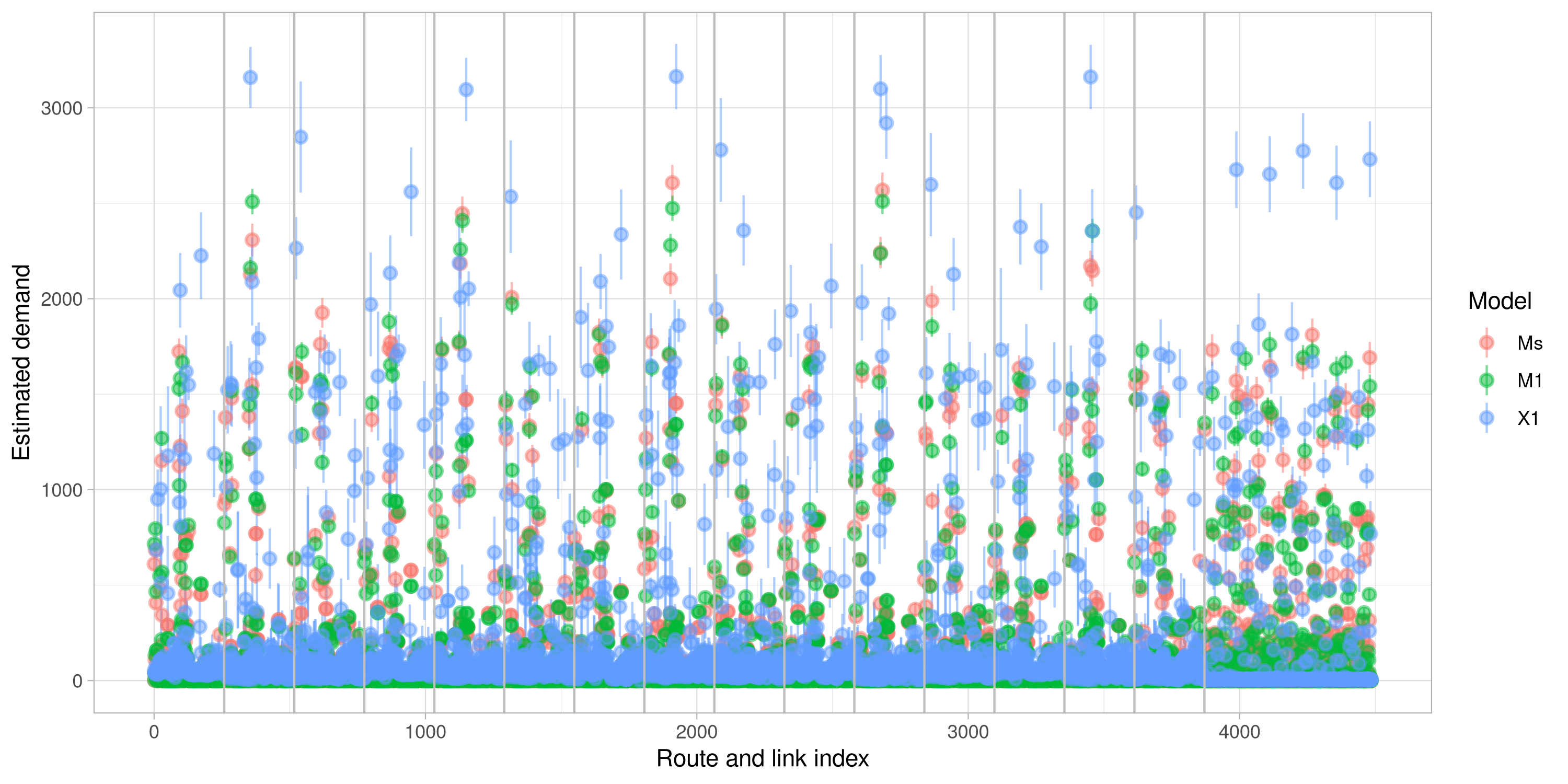}
\caption{Estimated demand for models $M_s$, $M_1$, and regularized model with
design $X_1$. The first indices cover routes for each date and time period of
the study, with lines separating day and within day periods (e.g., first
window is October 10th 2022, Monday, 6--7AM, followed by the 8--9AM period in
the same day and so on). The last window covers the link slack, interpreted as
extra route, demands.}
\label{fig:bo4mobMfits}
\end{figure}

As Figure~\ref{fig:bo4mobMfits} shows, the estimated route flows seem to be
similar for each weekday, which suggests a model that assumes that daily
patterns are the same,
\[
\log \Exp[Z_{rdt}] = \sum_{l \in L(r)} \mu_l + \tau_{tl} I(t > 1)
\quad \text{and} \quad
\log \Exp[S_{ldt}] = \mu_l + \tau_{tl} I(t > 1). \tag{$M_1$}
\]
An even simpler model, considering only multiplicative effects by date and
time period on $\Exp[Z_{rdt}]$ is given by
\[
\begin{split}
  \log \Exp[Z_{rdt}] &= \sum_{l \in L(r)} \mu_l +
  \delta_d I(d > 1) + \tau_t I(t > 1) \qquad \text{and} \\
  \log \Exp[S_{ldt}] &= \mu_l + \delta_d I(d > 1) + \tau_t I(t > 1).
\end{split} \tag{$M_0$}
\]
As expected from the design matrices and evidenced by the estimated
route flows in Figure~\ref{fig:bo4mobMfits}, a significant part of the
flows can be attributed to the link count slacks since they are related by the
same coefficients.

We summarize the model fits in Figure~\ref{fig:bo4modlmodels}, where
the base date-time contrast design $X_{DT}$ and time contrast design $X_T$ are
given by
\[
X_{DT} := [ \one_D \otimes \one_T \quad
I_{D,-1} \otimes \one_T \quad
\one_D \otimes I_{T,-1}]
\quad \text{and} \quad
X_T := [ \one_D \otimes \one_T \quad
\one_D \otimes I_{T,-1}].
\]
All models are too rigid, with a fixed link slack structure, and are strongly
rejected at any reasonable significance level. Model $M_0$ is clearly
discrepant, being too simple for us to entertain further. Qualitatively,
however, it seems acceptable to consider $M_1$ relative to $M_2$ if we allow
for a more flexible link slack design.

\begin{figure}[htbp]
\centering
\begin{tabular}{lllrr}
\toprule
Model & Design & Null hypothesis & Deviance & DF \\
\midrule
$M_2$ &
  $\begin{bmatrix}
    X_{DT} \otimes M_0 \\ X_{DT} \otimes I_{|\mathcal{L}|}
  \end{bmatrix}$ &
  -- & 72179.3 & 4198 \\
$M_1$ &
  $\begin{bmatrix}
    X_T \otimes M_0 \\ X_T \otimes I_{|\mathcal{L}|}
  \end{bmatrix}$ &
  $H_0: \delta_{d1} = \cdots = \delta_{d|\mathcal{L}|} = 0$ &
  86013.7 & 4362 \\
$M_0$ &
  $\begin{bmatrix}
    \one_{DT} \otimes M_0 \quad X_{DT,-1} \otimes
    \one_{|\mathcal{R}|} \\
    \one_{DT} \otimes I_{|\mathcal{L}|} \quad
    X_{DT,-1} \otimes \one_{|\mathcal{L}|}
  \end{bmatrix}$ &
  \makecell[l]{$H_0: \delta_{d1} = \cdots = \delta_{d|\mathcal{L}|},$\\
  $\qquad\tau_{t1} = \cdots = \tau_{t|\mathcal{L}|}$} &
  372299.5 & 4438 \\
\bottomrule
\end{tabular}
\caption{Likelihood-based models. The null hypotheses are relative to $M_2$
and span $d = 1, \ldots, D$ and $t = 1, \ldots, T$.}
\label{fig:bo4modlmodels}
\end{figure}

In the absence of predictors such as travel costs or more refined effects that
account for user decisions (as in stochastic user equilibrium schemes),
we then pursue a design that simply separates route and link slack effects;
because such a design is over-determined---the observed link counts saturate
the link slack effects---we resort to a more non-parametric approach using
regularization, as in Section~\ref{ssec:leicester}.
The main difference relative to the Leicester example are the link count
slacks, but we can regard them as extra OD routes, in the same way we
considered OD slacks in OD matrix estimation as coming from external zones.

We consider two models:
\[
\log \Exp[Z_{rdt}] = \zeta_{rdt} \qquad \text{and} \qquad
\log \Exp[S_{ldt}] = \gamma_{ldt} \tag{$X_1$}
\]
with design matrix
\[
X_1 = \begin{bmatrix}
I_{DT} \otimes I_{|\mathcal{R}|} & 0 \\
0 & I_{DT} \otimes I_{|\mathcal{L}|}
\end{bmatrix},
\]
and
\[
\log \Exp[Z_{rdt}] = \zeta_{rt} \qquad \text{and} \qquad
\log \Exp[S_{ldt}] = \gamma_{ldt} \tag{$X_2$}
\]
with the same route and time period effects across all date levels and
corresponding design matrix
\[
X_2 = \begin{bmatrix}
\one_D \otimes I_T \otimes I_{|\mathcal{R}|} & 0 \\
0 & I_{DT} \otimes I_{|\mathcal{L}|}
\end{bmatrix}.
\]
In both models there are separated route effects $\zeta$ and link slack
effects $\gamma$ and we take $\beta^\top = [\zeta^\top ~ \gamma^\top]$.

For the prior we first define weights using the Ochiai similarity
in~\eqref{eq:laplacianweights} and noting that a single observation link
counts as an extra route. This way, the weight matrix can be split into route
and link blocks,
\[
W = \begin{bmatrix} W_r & W_{lr}^\top \\ W_{lr} & W_l \end{bmatrix}.
\]
The weight matrices for the models with designs $X_1$ and $X_2$ have then the
route and link blocks replicated across date and time periods,
\[
W_1 = \begin{bmatrix}
I_{DT} \otimes W_r & I_{DT} \otimes W_{lr}^\top \\
I_{DT} \otimes W_{lr} & I_{DT} \otimes W_l
\end{bmatrix}
\quad \text{and} \quad
W_2 = \begin{bmatrix}
I_T \otimes W_r & \one_D^\top \otimes I_T \otimes W_{lr}^\top \\
\one_D \otimes I_T \otimes W_{lr} & I_{DT} \otimes W_l
\end{bmatrix},
\]
respectively. Finally, to define the Laplacian we scale the blocks by their
respective sizes by defining
\[
\Lambda_1 = \begin{bmatrix}
I_{DT|\mathcal{R}|} & 0 \\
0 & \frac{|\mathcal{R}|}{|\mathcal{L}|} I_{DT|\mathcal{L}|}
\end{bmatrix}
\quad \text{and} \quad
\Lambda_2 = \begin{bmatrix}
I_{T|\mathcal{R}|} & 0 \\
0 & \frac{|\mathcal{R}|}{|\mathcal{L}|} I_{DT|\mathcal{L}|}
\end{bmatrix}
\]
and setting $\Omega_1 = \Lambda_1^{1/2} W_1 \Lambda_1^{1/2}$ and
$\Omega_2 = \lambda \Lambda_2^{1/2} W_2 \Lambda_2^{1/2}$.
The priors are then
$\beta \sim N(0, \lambda^{-1}\Omega_1^-)$ for the model with design $X_1$ and 
$\beta \sim N(0, \lambda^{-1}\Omega_2^-)$ for the model with design $X_2$
for some hyper-prior parameter $\lambda$ acting as a regularization penalty.
Using leave-one-out cross-validation we set $\lambda = 0.5$.

\begin{figure}[htbp]
\centering
\includegraphics[width=.9\textwidth]{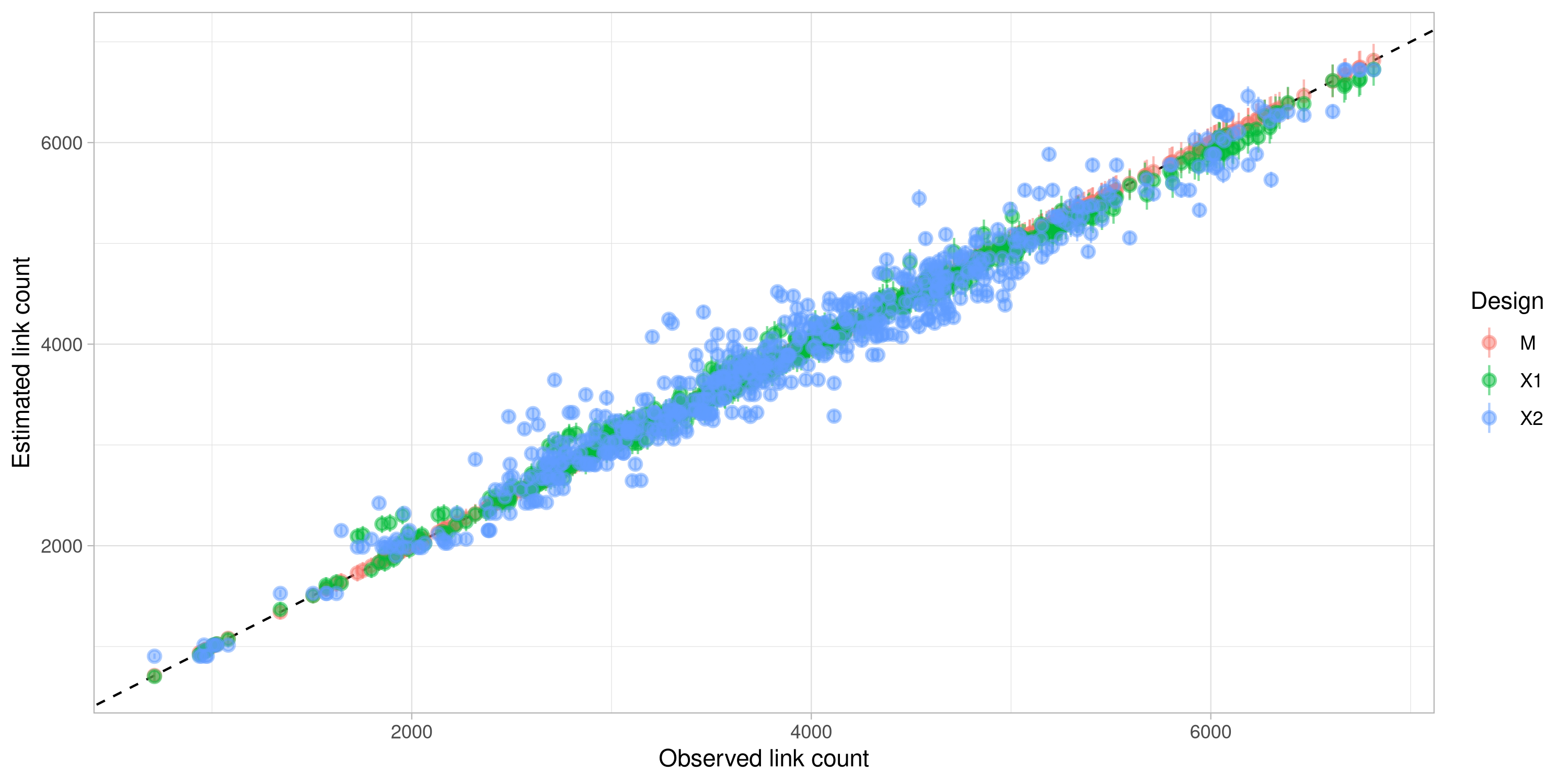}
\includegraphics[width=.9\textwidth]{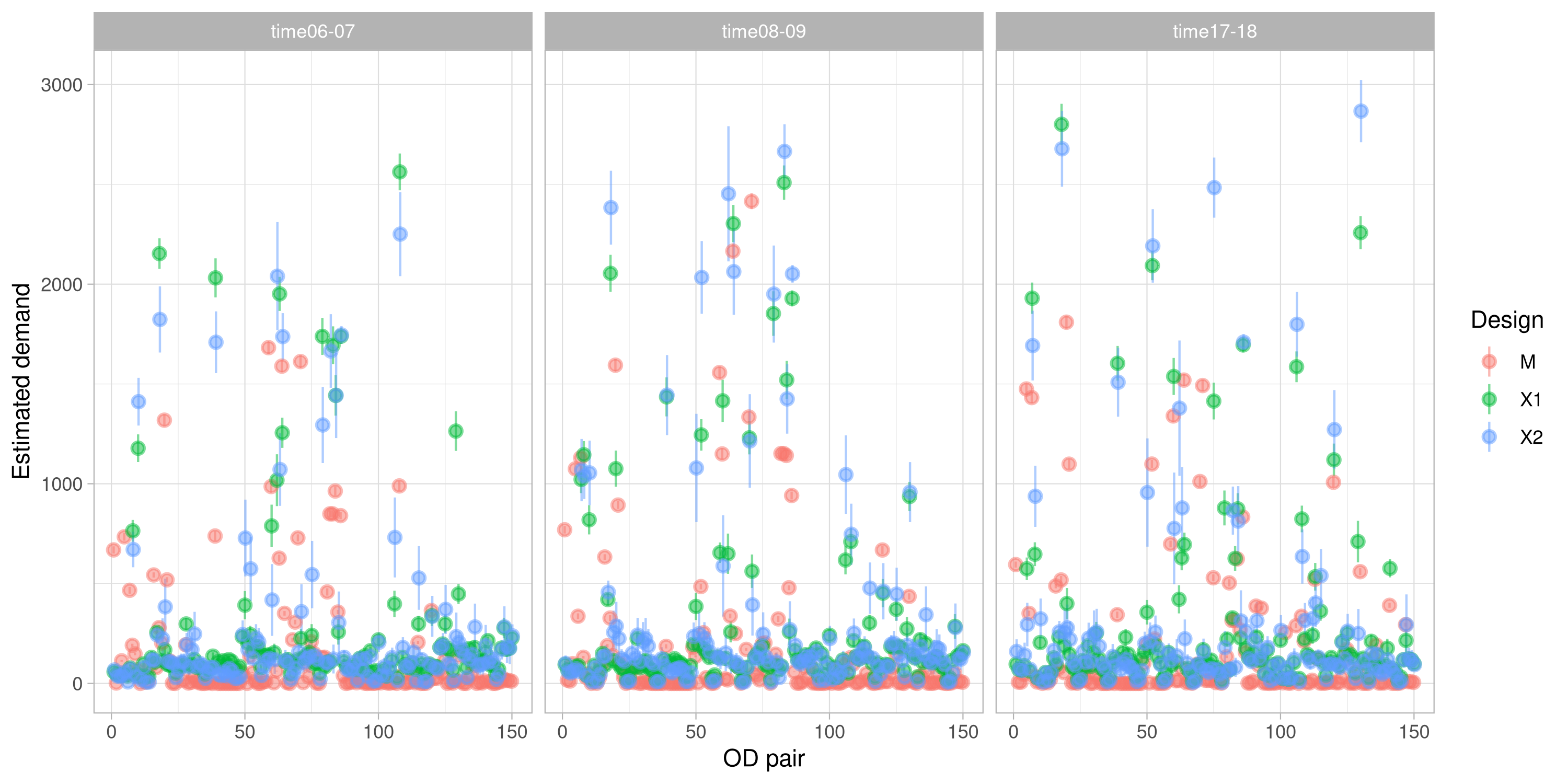}
\caption{BO4Mob estimated OD link counts (top) and OD route demands, averaged
by time period (bottom) for models $M_s$ (flat prior, with design labeled as
$M$) and regularized models with designs $X_1$ and $X_2$.}
\label{fig:bo4mobresults}
\end{figure}

The top panel of Figure~\ref{fig:bo4mobresults} assesses goodness of fit with
respect to observed link counts. Clearly, the saturated model $M_s$ is
linearly consistent, but we include it here as a sanity check and for
comparison of standard errors. The full model with design $X_1$ has overall
good fit with a few links showing regularization bias: higher estimates than
expected for lower count values and lower estimates than expected for higher
counts. The model with simpler, no day-to-day contrasts in the design $X_2$
shows worse fit with higher discrepancies, as expected.

\begin{figure}[p]
\centering
\includegraphics[width=.9\textwidth]{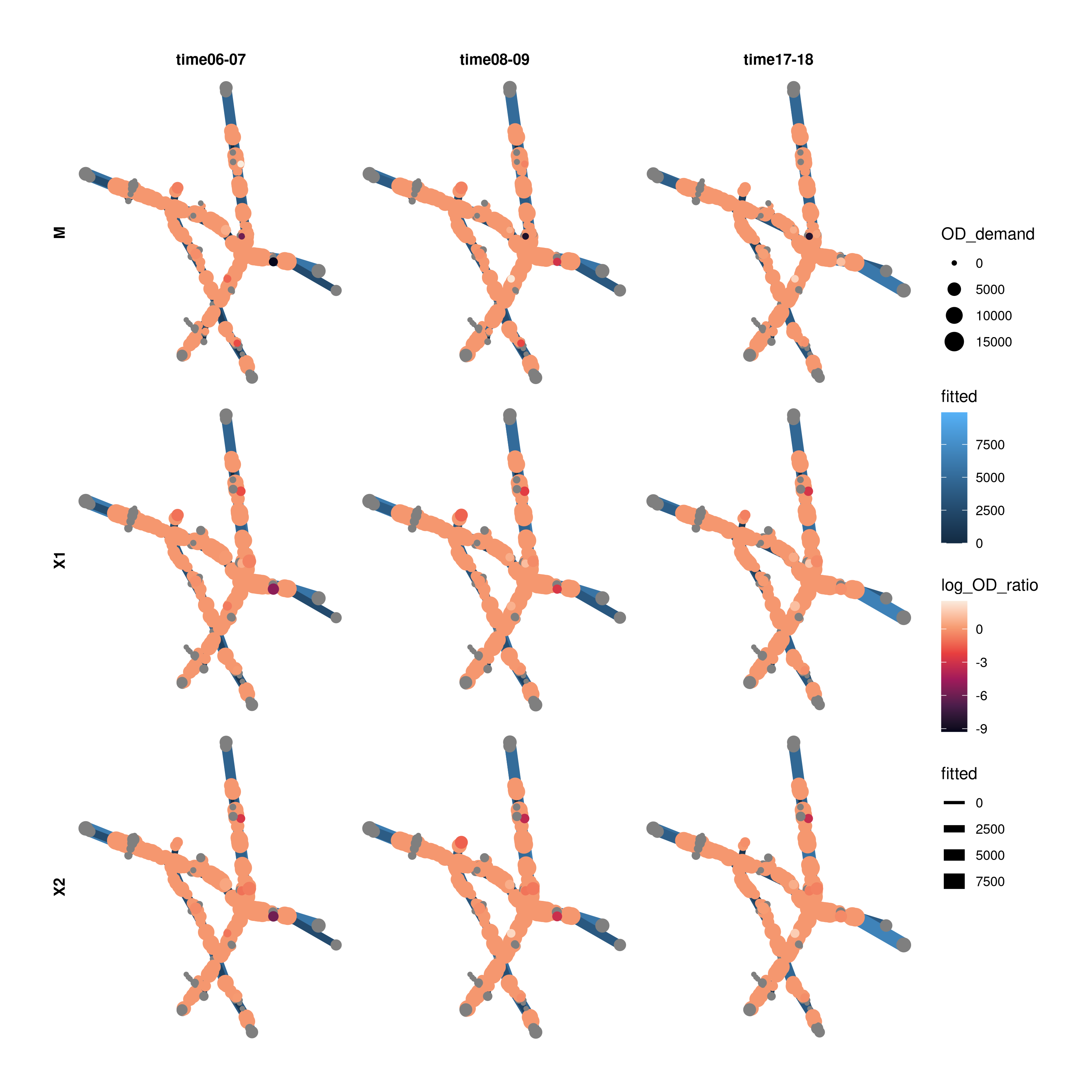}
\caption{BO4Mob route demands averaged by time period and for models with
designs $M$ (flat prior), $X_1$ and $X_2$ (regularized prior), loaded into
links of \texttt{4smallRegion} network. Network junctions are represented as
points with size proportional to total destination (in) and origin (out)
demand and by log origin-to-destination demand ratio.}
\label{fig:bo4mobnetwork}
\end{figure}

The lower panel of Figure~\ref{fig:bo4mobresults} summarizes the time period
demand patterns by taking averaged estimated demands for each period. Because
$M_s$ has higher estimated link slacks, we have lower OD demand estimates for
the routes. The regularized models, on the other hand, more effectively shrink
link slack estimates, as can also be seen in Figure~\ref{fig:bo4mobMfits}.
High link slack estimates in the model with design $X_1$ at
Figure~\ref{fig:bo4mobMfits}, especially around 2700 for a single link in the
morning peak across all study dates, suggest the need for a better coverage of
routes by measurement links.
The model with design $X_2$ has higher OD estimates for some selected routes
and larger standard errors due to lower regularization since there are fewer
route demand terms. Qualitatively, the loaded demand on network links are
similar across the models, as seen in Figure~\ref{fig:bo4mobnetwork}.

\section{Conclusion}
We have shown that data censored by linear constraints can be fit within the
generalized linear model framework, turning a problem usually handled by
optimization or expensive sampling into a standard Fisher score based GLM
procedure. This is achieved by considering restricted curved exponential
families and avoiding marginalization of latent observations.
The payoff is that estimation and uncertainty quantification come
together: alongside the reconstructed OD demands we obtain coefficient standard
errors, residuals, and a dispersion estimate, at a fraction of the cost of
MCMC. Across the synthetic studies the method recovered the diagonal-dominant,
non-dominant, and cost-constrained structures on demand through its priors and
design extensions. In particular, when applied to network tomography even
without reliable predictors such as travel costs and utilities, Laplacian
priors offer a regularized way of characterizing route and OD demands from
observed link flows, allowing for descriptive non-parametric estimates of
large transit systems. We hope that this formulation enables more
computationally efficient procedures that better integrate with other parts of
an urban planning system, such as in the trip distribution and allocation
stages of a larger, encompassing four-step transportation model.

A number of extensions seem promising. The first is a fully Bayesian treatment
that characterizes the posterior space of coefficients via Hamiltonian Monte
Carlo sampling~\citep{girolami2011riemann}. We also intend to allow for
dispersed families, e.g. negative binomial, by developing sampling procedures
to estimate dispersion parameters based on extended quasi-likelihood
approaches~\citep{nelder1987quasi}.
Sampling the latent counts $Z$ directly, that is, reconstructing the flows,
has been recently proposed by Hazelton, and we plan to incorporate this class
of samplers in our methodology using a collapsed sampler.
Finally, we plan to explore our methodology to migration studies, where the
same margin-censoring structure appears but with an added birthplace dimension
that introduces cross-table constraints; adapting the constraint and design
matrices to that three-dimensional setting is the subject of ongoing work.


\newpage

\bibliography{ref}


\end{document}